\documentclass[prd,aps,eqsecnum,amsmath,amssymb,floatfix,nofootinbib,preprint,preprintnumbers,tightenlines]{revtex4}

\usepackage{bm}
\usepackage{epsfig}
\usepackage{float}
\usepackage{color}
\let\ifcomments\iftrue
\def\commentsoff{\global\let\ifcomments\iffalse}

\let\commentsize\small
\def\tinycomments{\global\let\commentsize\footnotesize}

\begin{document}
\font\rm=cmr12
\font\bf=cmbx12
\font\it=cmti12
\rm

\thispagestyle{empty}

\begin{center}
\vspace{10mm}
\begin{boldmath}
{\large\bf Evaluation of the three-flavor quark-disconnected 
contribution\\ 
to the muon anomalous magnetic moment from experimental data}
\end{boldmath}

\vspace{3ex}
{Diogo~Boito,$^{a,b}$ Maarten~Golterman,$^{c,d}$
 Kim~Maltman,$^{e,f}$ and  Santiago~Peris$^d$%
\\[0.1cm]
{\it
\null$^a$Instituto de F{\'i}sica de S{\~a}o Carlos, Universidade de 
S{\~a}o Paulo,\\ CP 369, 13570-970, S{\~a}o Carlos, SP, Brazil\\
\null$^b$University of Vienna, Faculty of Physics, Boltzmanngasse 5,
A-1090 Wien, Austria \\
\null$^c$Department of Physics and Astronomy\\
San Francisco State University, San Francisco, CA 94132, USA\\
\null$^d$Department of Physics and IFAE-BIST, Universitat Aut\`onoma 
de Barcelona,\\ E-08193 Bellaterra, Barcelona, Spain\\
\null$^e$Department of Mathematics and Statistics\\
York University,  Toronto, ON Canada M3J~1P3\\
\null$^f$CSSM, University of Adelaide, Adelaide, SA~5005 Australia}
\\[6mm]}

\end{center}

\begin{abstract}
\vskip0.5cm
We point out that the sum of the strange-quark-connected and full,
three-flavor quark-disconnected contributions to the leading-order 
hadronic vacuum polarization contribution, $a_\mu^{\mathrm{LO,HVP}}$, to
the anomalous magnetic moment of the muon, is a physical observable, and 
provide a data-based determination of this quantity in the isospin limit. 
The result, $40.1(1.5)\times 10^{-10}$, or $38.7(2.0)\times 10^{-10}$, 
depending on which data compilation is used, serves as a target of 
comparison for lattice calculations of the same isospin-limit combination. 
Subtracting from this result the average of lattice determinations of the 
strange-quark-connected contribution, one obtains also an alternate 
determination of the isospin-limit three-flavor disconnected contribution 
to $a_\mu^{\mathrm{LO,HVP}}$. The result of this determination, 
$-13.3(1.5)\times 10^{-10}$, or $-14.6(2.0)\times 10^{-10}$, 
depending on which data compilation is used, agrees well and 
is competitive with the most precise current lattice determination. 
\end{abstract}

\preprint{\it University of Vienna preprint UWThPh 2022-4}
\maketitle

\section{Introduction}\label{intro}
Interest in reducing the uncertainty on the Standard Model (SM) 
prediction for $a_\mu$, the anomalous magnetic moment of the muon,
increased dramatically with the release of the 2006 BNL E821 
experimental result~\cite{Muong-2:2006rrc}, which showed a more 
than $3.5\sigma$ tension with then-existing SM expectations. This 
interest has been further heightened by the release of the first results 
from the Fermilab E989 experiment~\cite{Muong-2:2021ojo}, which raise the 
discrepancy between the SM expectation and the experimental world
average to the $4.2\sigma$ level. The main source of uncertainty in 
the SM prediction is currently that on hadronic contributions, in 
particular the leading-order hadronic vacuum polarization (LO, HVP) 
contribution, $a_\mu^{\mathrm{LO,HVP}}$. While a dispersive evaluation of 
this contribution is possible using experimental hadronic cross-section 
data, the result, in fact, represents the SM expectation for this 
quantity only if beyond-the-SM contributions to the appropriately
weighted integral of the experimental cross sections are numerically 
negligible. While this is likely the case at the current level of 
precision, a first-principles lattice determination of the SM expectation 
for $a_\mu^{\mathrm{LO,HVP}}$ is of interest in its own right, and 
there has been intense recent activity aimed at reduced the uncertainty 
on such lattice determinations~\cite{Chakraborty:2014mwa,RBC:2015you,Chakraborty:2016mwy,RBCUKQCD:2016clu,Giusti:2017ier,FermilabLattice:2017wgj,Budapest-Marseille-Wuppertal:2017okr,Giusti:2017jof,RBC:2018dos,Giusti:2018mdh,Giusti:2018vrc,Gulpers:2018mim,Giusti:2019xct,Shintani:2019wai,FermilabLattice:2019ugu,Gerardin:2019rua,Aubin:2019usy,Borsanyi:2020mff,Lehner:2020crt,Aubin:2021vej,Giusti:2021dvd,Risch:2021hty},
with the most recent BMW result~\cite{Borsanyi:2020mff} reaching, for the
first time in a lattice determination, the sub-percent precision level. 
First-principles lattice determinations also avoid experimental issues 
present in the alternate dispersive determination, such as the impact of 
the long-standing inconsistencies between BaBar and KLOE results for the 
$e^+ e^-\rightarrow \pi^+ \pi^-$ cross sections. 

Lattice determinations of $a_\mu^{\mathrm{LO,HVP}}$ typically involve 
evaluating and summing isospin-limit light-, strange-, charm- and 
bottom-quark connected contributions, the isospin-limit disconnected 
contribution, and both electromagnetic (EM) and strong isospin breaking 
(SIB) contributions. {\footnote{{By ``isospin limit," we will 
always mean pure QCD with no EM corrections and $m_d=m_u$.}}}
The lattice calculation of the disconnected contribution is particularly
numerically intensive, and only a limited number of lattice results
exist for this quantity. 

Comparisons between the results from different lattice collaborations 
for each of the individual flavor-specific connected contributions, 
the disconnected contribution, and the EM and SIB contributions are 
useful for identifying and controlling lattice systematic effects, as 
are comparisons between results from different groups for related 
quantities such as the intermediate-Euclidean-time-window integrals, 
$a_\mu^W$, introduced by RBC/UKQCD~\cite{RBC:2018dos}. The window 
quantities are also of interest since they have alternate dispersive 
representations and hence allow comparisons of lattice and dispersive, 
data-based results for contributions to $a_\mu^{\mathrm{LO,HVP}}$ from 
different windows in Euclidean time. Additional observables amenable to 
both lattice and dispersive determinations, ideally with relative $I=1$ 
and $I=0$ contributions different from those of the RBC/UKQCD window 
quantities, would be of interest for a similar reason.

In this paper, we point out that the sum of strange-quark-connected 
and full, three-flavor disconnected contributions to 
$a_\mu^{\mathrm{LO,HVP}}$, denoted $a_\mu^{\mathrm{sconn+disc}}$ in 
what follows, provides an example of such an additional observable, 
one with a significantly higher relative weight for $I=0$ contributions. 
We then show how existing experimental results can be used to obtain a 
data-based determination of $a_\mu^{\mathrm{sconn+disc}}$. The precision 
on this determination turns out to be such that, using existing lattice 
results for the strange-quark-connected contribution, one is able to 
determine the three-flavor disconnected contribution to 
$a_\mu^{\mathrm{LO,HVP}}$ with a precision comparable to that of the 
best current lattice determination.

The rest of the paper is organized as follows. Sec.~\ref{notation} lays 
out some relevant background and notation, while Sec.~\ref{basicidea} 
outlines the basic analysis approach. Sec.~\ref{knt19det} provides details 
of a numerical implementation using as input (i) the results of the BaBar
determination of the differential $\tau\rightarrow K^- K^0\nu_\tau$ 
decay distribution~\cite{BaBar:2018qry} and (ii) the assessment of 
exclusive-mode $e^+e^-\rightarrow hadrons$ cross sections 
and exclusive-mode contributions to $a_\mu^{\mathrm{LO,HVP}}$ detailed in 
Refs.~\cite{Keshavarzi:2018mgv,Keshavarzi:2019abf}. In Sec.~\ref{ibcorrns}
we discuss isospin-breaking corrections to the results obtained in
Sec.~\ref{knt19det}, needed to make contact with isospin-symmetric lattice 
results. Sec.~\ref{dhmz} outlines an alternate version of the full 
analysis employing, in place of the exclusive-mode $a_\mu^{\mathrm{LO,HVP}}$ 
contributions of Refs.~\cite{Keshavarzi:2018mgv,Keshavarzi:2019abf},
the alternate set of such contributions from Ref.~\cite{Davier:2019can}.
Finally, in Sec.~\ref{discussion}, we summarize and briefly discuss our 
results. In the rest of the paper we will, for the sake of brevity,
use the shorter forms ``disconnected contribution'' and 
``strange-quark-connected plus disconnected contribution'' in place of the 
more accurate, but longer, expressions ``full, three-flavor disconnected 
contribution'' and ``strange-quark-connected plus full, three-flavor 
disconnected contribution''.

\section{Notation and background}\label{notation}
In this section, we define our notation and introduce a number of useful 
decompositions.  

We denote the flavor octet of light- and strange-quark vector currents as
$V^a_\mu =\bar{q}{\frac{\lambda^a}{2}}\gamma_\mu q$, $a=1,\cdots 8$, 
with $q$ the column vector $(u,d,s)^T$.  This allows us to define the
vector current two-point functions, $ \Pi^{ab}_{\mu\nu}(q)$, their
associated polarizations, $\Pi^{ab}(Q^2)$, and the associated spectral 
functions, $\rho^{ab}(s)$, with $s=q^2$ and $Q^2\, =\, -q^2$, 
as usual, by
\begin{eqnarray}
&&\Pi^{ab}_{\mu\nu}(q) = (q_\mu q_\nu - q^2 g_{\mu\nu})\Pi^{ab}(Q^2) =
i\, \int d^4 x e^{iq\cdot x} \langle 0 \vert 
T \left( V^a_\mu (x) V^b_\nu (0) \right)\vert 0\rangle \, , \label{poln}\\
&&\rho^{ab}(s)={\frac{1}{\pi}}\, \mbox{Im}\, \Pi^{ab}(Q^2)\ ,\quad
(s=\, -Q^2>0)\ ,\label{specfunc}
\end{eqnarray}
and the subtracted polarizations, $\hat{\Pi}^{ab}(Q^2)$, by
\begin{equation}
\hat{\Pi}^{ab}(Q^2)=\Pi^{ab}(Q^2)-\Pi^{ab}(0)\, .
\label{subdpoln}\end{equation}

The ($u,d,s$) part of the electromagnetic (EM) current, 
$J_\mu^{\mathrm{EM}}$, has the decomposition
\begin{equation}
J_\mu^{\mathrm{EM}} = V_\mu^3 + {\frac{1}{\sqrt{3}}} V_\mu^8 \,\equiv\,
J_\mu^{\mathrm{EM},3}+J_\mu^{\mathrm{EM},8}\, =\, 
{\frac{1}{2}}\left( \bar{u}\gamma_\mu u -\bar{d} \gamma_\mu d\right)
\, +\, {\frac{1}{6}}\left( \bar{u}\gamma_\mu u +\bar{d} \gamma_\mu d
-2\bar{s}\gamma_\mu s\right)
\label{emisodecomp}\end{equation}
into $I=1$ ($a=3$) and $I=0$ ($a=8$) parts. We also define, for use below, 
the coefficients $c_k^a$, $k=u,d,s$ and $a=3,8$, via
\begin{equation}
J_\mu^{\mathrm{EM},a} \equiv \sum_{k=u,d,s}c_k^a\, \bar{q}\gamma_\mu q \, .
\end{equation}
The values are $c_u^3\, =\, -c_d^3\, =\, 1/2$, $c^3_s=0$, and
$c_u^8\, =\, c_d^8\, =\, 1/6$, $c^8_s\, =\, -2/6$.

The subtracted three-flavor EM vacuum polarization and associated 
spectral function, $\rho_{\mathrm{EM}}(s)$, have the related decompositions,
\begin{eqnarray}
\hat{\Pi}_{\mathrm{EM}}(Q^2)&=&\hat{\Pi}_{\mathrm{EM}}^{33}(Q^2)
+{\frac{2}{\sqrt{3}}}
\hat{\Pi}_{\mathrm{EM}}^{38}(Q^2)+{\frac{1}{3}}
\hat{\Pi}_{\mathrm{EM}}^{88}(Q^2)\nonumber\\
&\equiv&  \hat{\Pi}_{\mathrm{EM}}^{I=1}(Q^2)
+\hat{\Pi}_{\mathrm{EM}}^{\mathrm{MI}}(Q^2)+
\hat{\Pi}_{\mathrm{EM}}^{I=0}(Q^2)\ ,\nonumber\\
\rho_{\mathrm{EM}}(s)&=&\rho^{33}(s)+{\frac{2}{\sqrt{3}}}\rho^{38}(s)+
{\frac{1}{3}}\rho^{88}(s)\nonumber\\
& \equiv & \rho_{\mathrm{EM}}^{I=1}(s)+
\rho_{\mathrm{EM}}^{\mathrm{MI}}(s)+\rho_{\mathrm{EM}}^{I=0}(s)\, ,
\label{piemrhoemdecomps}\end{eqnarray}
into pure isovector ($ab=33$, $I=1$), pure isoscalar ($ab=88$, $I=0$), and 
mixed isospin ($ab=38,83$, $MI$) parts, with the latter, of course, vanishing 
in the isospin limit. In the isospin limit, $\hat{\Pi}^{33}$ has only 
light-quark-connected contributions, while $\hat{\Pi}^{88}$ is a sum of 
light-quark-connected, strange-quark-connected and all disconnected 
contributions. 

The hadronic contribution $a_\mu^{\mathrm{LO,HVP}}$ can be determined 
using the standard ``dispersive'' representation,
\begin{equation}
a_\mu^{\mathrm{LO,HVP}}={\frac{\alpha_{\mathrm{EM}}^2m_\mu^2}{9\pi^2}}
\int_{m_\pi^2}^\infty ds\, {\frac{\hat{K}(s)}{s^2}} R(s)\, ,
\label{amudispform}\end{equation}
where $\alpha_{\mathrm{EM}}$ is the EM fine-structure constant, $R(s)$ is the
standard EM cross-section ratio,
\begin{equation}
R(s)={\frac{3s}{4\pi \alpha_{\mathrm{EM}}^2}}\, 
\sigma^{(0)} [e^+ e^-\rightarrow hadrons(+\gamma)]\, ,
\label{rdefn}\end{equation}
with $\sigma^{(0)} [e^+ e^-\rightarrow hadrons(+\gamma)]$ the bare
inclusive hadronic electroproduction cross section. The kernel $\hat{K}(s)$
is exactly known and slowly (and monotonically) increasing with $s$ (see 
for example Ref.~\cite{Aoyama:2020ynm}). The dispersive determination 
typically employs a sum of exclusive-mode contributions up to just below 
$s=4$ GeV$^2$, and inclusive $R(s)$ determinations and/or perturbative QCD 
(pQCD) above that, apart from in the region of narrow charm and bottom 
resonances. Using the relation
\begin{equation}
R(s)=12\pi^2 \rho_{\mathrm{EM}}(s)\ ,
\label{Rsintermsofrhoem}
\end{equation}
The three-flavor contribution to
$a_\mu^{\mathrm{LO,HVP}}$ can also be broken down into $I=1$ ($33$), $I=0$ 
($88$) and mixed-isospin ($MI$, $38+83$) contributions,
\begin{equation}
a_\mu^{\mathrm{LO,HVP}} = a_\mu^{33}+{\frac{2}{\sqrt{3}}}a_\mu^{38}
+{\frac{1}{3}} a_\mu^{88}\equiv a_\mu^{I=1}+a_\mu^{\mathrm{MI}}+a_\mu^{I=0}\, .
\label{amulohvpbreakdown}\end{equation}
Contributions to these quantities from an individual
exclusive mode, $X$, can also be defined, and are denoted
$[a_\mu^{\mathrm{LO,HVP}}]_X$, $[a_\mu^{I=1}]_X$, $[a_\mu^{I=0}]_X$ and 
$[a_\mu^{\mathrm{MI}}]_X$. 

The hadronic contribution $a_\mu^{\mathrm{LO,HVP}}$ also has the standard 
weighted Euclidean-$Q^2$ integral 
representation~\cite{Lautrup:1971jf,deRafael:1993za,Blum:2002ii},
\begin{equation}
a_\mu^{\mathrm{LO,HVP}} =\, -4\alpha_{\mathrm{EM}}^2 \int_0^\infty dQ^2 f(Q^2) 
\hat{\Pi}_{\mathrm{EM}}(Q^2)\, ,
\label{amueuclhvplatt}\end{equation}
with $f(Q^2)$ another exactly known kernel which diverges as $1/\sqrt{Q^2}$ 
as $Q^2\rightarrow 0$ and creates a peak in the integrand of 
Eq.~(\ref{amueuclhvplatt}) at very low $Q^2\simeq m_\mu^2/4$. This expression,
or the related time-momentum representation~\cite{Bernecker:2011gh}, forms 
the basis for lattice determinations of $a_\mu^{\mathrm{LO,HVP}}$. Analogous
representations for $a_\mu^{I=1}$, $a_\mu^{I=0}$ and $a_\mu^{\mathrm{MI}}$
are obtained by replacing $\hat{\Pi}_{\mathrm{EM}}$ in 
Eq.~(\ref{amueuclhvplatt}) with 
$\hat{\Pi}_{\mathrm{EM}}^{I=1}$, $\hat{\Pi}_{\mathrm{EM}}^{I=0}$ and 
$\hat{\Pi}_{\mathrm{EM}}^{\mathrm{MI}}$, respectively.
To first order in $m_d-m_u$ there are no SIB contributions to either 
$a_\mu^{33}$ or $a_\mu^{88}$, while SIB is expected to dominate $a_\mu^{38}$. 

\section{The basic idea}\label{basicidea}
The basic idea of the analysis is the following.
In the isospin limit, $\hat{\Pi}_{\mathrm{EM}}^{I=1}$ receives only
light-quark-connected contributions, while $\hat{\Pi}_{\mathrm{EM}}^{I=0}$ 
is a sum of light-quark-connected, strange-quark-connected and all 
disconnected contributions. It is thus obvious that there exists a 
combination of $\hat{\Pi}_{\mathrm{EM}}^{I=1}$ and 
$\hat{\Pi}_{\mathrm{EM}}^{I=0}$ in which the light-quark-connected 
contributions cancel, leaving a result which is the sum of the 
strange-quark-connected and disconnected contributions. 
It is easily checked (as noted explicitly in Ref.~\cite{RBC:2015you})
that this combination is
\begin{equation}
\hat{\Pi}_{\mathrm{EM}}^{\mathrm{sconn+disc}}\equiv 
\hat{\Pi}_{\mathrm{EM}}^{I=0}\, -\, 
{\frac{1}{9}}\, \hat{\Pi}_{\mathrm{EM}}^{I=1}\ .
\label{strangeconnplusdisc}\end{equation}
The corresponding spectral function is
\begin{equation}
\rho_{\mathrm{EM}}^{\mathrm{sconn+disc}}(s)=\rho_{\mathrm{EM}}^{I=0}(s)\, 
-\, {\frac{1}{9}}\, \rho_{\mathrm{EM}}^{I=1}(s)\ .
\label{sconnplusdiscspecfunc}\end{equation}
The appropriately weighted dispersive integral of this latter combination
produces a result, $a_\mu^{\mathrm{sconn+disc}}$, which is the sum of the 
strange-quark-connected and disconnected contributions to 
$a_\mu^{\mathrm{LO,HVP}}$. It follows that, if the $I=0$ and $I=1$ 
contributions to $R(s)$ can be  separated with sufficient precision, an 
accurate experimental determination of this combination will be possible. In 
terms of the $I=0$ and $I=1$ contributions to $a_\mu^{\mathrm{LO,HVP}}$, 
\begin{equation}
a_\mu^{\mathrm{sconn+disc}}=a_\mu^{I=0}\, -\, {\frac{1}{9}}\, a_\mu^{I=1}\, .
\label{nominalsconnpdisc}\end{equation}
Such a determination would serve as a useful target of comparison for lattice 
determinations of this same sum.  

The strange-quark-connected contribution to $a_\mu^{\mathrm{LO,HVP}}$, 
$a_\mu^{\mathrm{sconn}}$, has been rather precisely determined by 
several lattice groups~\cite{Chakraborty:2014mwa,Budapest-Marseille-Wuppertal:2017okr,RBCUKQCD:2016clu,RBC:2018dos,Shintani:2019wai,Gerardin:2019rua,Giusti:2018mdh,Borsanyi:2020mff}. 
The most recent (BMW) determination,
$53.393(89)(68)\times 10^{-10}$~\cite{Borsanyi:2020mff}, is in excellent
agreement with the average, $(53.2\pm 0.3)\times 10^{-10}$, of previous 
determinations quoted in the 2020 $g-2$ Theory Initiative
white paper~\cite{Aoyama:2020ynm}. $a_\mu^{\mathrm{sconn}}$ is thus already 
known to much higher precision than the final target $\sim 1.4\times 10^{-10}$
uncertainty on $a_\mu$ expected from the full FNAL E989 experimental program. 
In view of this precision, an experimental determination of 
$a_\mu^{\mathrm{sconn+disc}}$ will also provide a determination, with 
comparable precision, of the full disconnected contribution to 
$a_\mu^{\mathrm{LO,HVP}}$, $a_\mu^{\mathrm{disc}}=a_\mu^{\mathrm{sconn+disc}}
-a_\mu^{\mathrm{sconn}}$. All this, of course, assumes the 
experimentally determined $a_\mu^{\mathrm{sconn+disc}}$ combination 
does not contain significant beyond-the-SM contributions and 
hence should be compatible with SM-based lattice determinations
of this quantity.

In Sec.~\ref{knt19det} we will implement this idea neglecting,
to begin with, isospin-breaking (IB) corrections. Then, in 
Sec.~\ref{ibcorrns} we will take into account IB corrections to 
arrive at our final results for $a_\mu^{\mathrm{sconn+disc}}$ and 
$a_\mu^{\mathrm{disc}}$.

\section{An implementation with current experimental data}\label{knt19det}
In what follows, we employ the dispersive results for exclusive-mode 
contributions to $a_\mu^{\mathrm{LO,HVP}}$ from the region 
$\sqrt{s}\leq 1.937$~ GeV, listed in Table 1 of 
Ref.~\cite{Keshavarzi:2019abf} (which we will refer to as KNT2019). 

Above $s=1.937^2$~GeV$^2$, we use the five-loop, $n_f=3$ pQCD expression 
for $\rho_{\mathrm{EM}}^{\mathrm{sconn+disc}}(s)$, with PDG2020 input for 
$\alpha_s$~\cite{ParticleDataGroup:2020ssz}. It is straightforward to show
that the pQCD result for $\rho_{\mathrm{EM}}^{\mathrm{sconn+disc}}(s)$ is 
one-sixth that for $\rho_{\mathrm{EM}}(s)$, in the chiral limit, with
small corrections proportional to $m_s^2/s$. The approximation of using the 
pQCD representation for $\rho_{\mathrm{EM}}^{\mathrm{\mathrm{sconn+disc}}}(s)$
in this region is expected to be an accurate one, up to possible small 
duality violating corrections. As illustrated in Fig.~12 of 
Ref.~\cite{Aoyama:2020ynm}, $n_f=3$ pQCD expectations for $R(s)$ agree 
within errors with experimental 
determinations~\cite{BES:2001ckj,BES:2009ejh,KEDR:2018hhr} (especially 
those of KEDR~\cite{KEDR:2018hhr}) in the region from slightly below 
$\sqrt{s}=2$~GeV up to the charm threshold.

The main part of the $I=1$/$I=0$ separation of exclusive-mode contributions
is accomplished, as usual, using $G$-parity. Exclusive modes with
positive/negative $G$-parity have $I=1$/$I=0$. This allows unique isospin
assignments for contributions from exclusive modes consisting entirely of 
strong-interaction-stable and/or narrow states with well-defined $G$-parity 
($\pi$, $\eta$, $\omega$, $\phi$). Such modes account for more than $93\%$ 
of the contribution to $a_\mu^{\mathrm{LO,HVP}}$ from the KNT2019 
exclusive-mode region. 

Additional information is required to separate the $I=1$ and $I=0$ components 
of the contributions of exclusive modes containing at least one $K\bar{K}$ 
pair, which are not eigenstates of $G$-parity. We discuss below how this 
separation can be accomplished using experimental input in the case of
the $K\bar{K}$ and $K\bar{K}\pi$ exclusive modes. For all remaining
KNT2019 $G$-parity-ambiguous exclusive modes, $X$, whose spectral 
contributions lie at higher $s$ and whose contributions to 
$a_\mu^{\mathrm{LO,HVP}}$ are thus strongly numerically suppressed, we 
use a ``maximally conservative'' assessment of the desired 
$[a_\mu^{\mathrm{sconn+disc}}]_X = [a_\mu^{I=0}]_X\,
-\,{\frac{1}{9}}\,[a_\mu^{I=1}]_X$ combination, determined as follows. 
Since the $I=0$ contribution for mode $X$ can, in principle, lie 
anywhere between $0$ and the full mode-$X$ $I=0+1$ total, 
$[a_\mu^{\mathrm{LO,HVP}}]_X$, the $[a_\mu^{\mathrm{sconn+disc}}]_X$ 
combination we are interested in necessarily lies between 
$-{\frac{1}{9}}[a_\mu^{\mathrm{LO,HVP}}]_X$ and 
$[a_\mu^{\mathrm{LO,HVP}}]_X$.{\footnote{Explicitly: if $x$ is the fraction 
of the total that is $I=0$, the $I=1$ fraction is $1-x$, and the 
$(I=0)-(1/9)(I=1)$ combination a fraction $x-(1/9)(1-x) = (10/9)x - (1/9)$ 
of the total. This result is monotonic in $x$, increasing from $-1/9$ at 
$x=0$ to $1$ at $x=1$.}} One may thus be maximally conservative and cover
this entire range by taking
\begin{equation}
[a_\mu^{\mathrm{sconn+disc}}]_X \, =\, 
\left( {\frac{4}{9}}\pm {\frac{5}{9}}\right)
\, [a_\mu^{\mathrm{LO,HVP}}]_X\ .
\end{equation}

We now turn to the explicit numerical determination of the 
strange-quark-connected plus full disconnected sum, using the data input and 
analysis strategies outlined above and described in more detail below. Note 
that the KNT2019 results used below for all $G$-parity-unambigous 
exclusive-mode contributions are the contributions of these modes between 
threshold and $\sqrt{s}=1.937$~GeV. Inclusive input is used above this point. 
In what follows, we outline how additional experimental input can be used to 
fix the $I=1$ contributions, $[\rho_{\mathrm{EM}}^{I=1}(s)]_X$, to 
$\rho_{\mathrm{EM}}(s)$, and hence also the $I=1$ contributions 
$[a_\mu^{I=1}]_X$, for the exclusive modes $X=K\bar{K}$ and $K\bar{K}\pi$. 
By carrying out the $[\rho_{\mathrm{EM}}^{I=1}(s)]_X$ determinations over 
the full KNT2019 exclusive-mode region, $\sqrt{s}\leq 1.937$~GeV, the 
associated $I=0$ contributions, and hence also the desired 
$[a_\mu^{\mathrm{sconn+disc}}]_X$ combinations, for that same region 
follow immediately from the corresponding KNT2019 total $I=0+1$ 
$[a_\mu^{\mathrm{LO,HVP}}]_X$ results.

\begin{table}[t]
\begin{center}
\begin{tabular}{lr|lr}
\hline
$I=1$ modes $X$&$[a_\mu^{\mathrm{LO,HVP}}]_X\times 10^{10}$&$I=0$ modes $X$&
$[a_\mu^{\mathrm{LO,HVP}}]_X\times 10^{10}$\\
\hline
low-$s$ $\pi^+ \pi^-$& 0.87(02)\quad& low-$s$ $3\pi$& 0.01(00)\quad\\
$\pi^+ \pi^-$& 503.46(1.91)\quad& $\pi^0\gamma$ ($\omega$, $\phi$ dominated)& 
4.46(10)\quad\\
$2\pi^+ 2\pi^-$& 14.87(20)\quad& $3\pi$& 46.73(94)\quad\\
$\pi^+ \pi^- 2\pi^0$& 19.39(78)\quad& $2\pi^+2\pi^-\pi^0$ (no $\omega$, 
$\eta$)& 0.98(09)\quad\\
$3\pi^+ 3\pi^-$ (no $\omega$)& 0.23(01)\quad& $\pi^+ \pi^- 3\pi^0$ (no $\eta$)&
0.62(11)\quad\\
$2\pi^+2\pi^-2\pi^0$ (no $\eta$)& 1.35(17)\quad& $3\pi^+3\pi^- \pi^0$ (no 
$\omega$, $\eta$)& 0.00(01)\quad\\
$\pi^+\pi^- 4\pi^0$ (no $\eta$)& 0.21(21)\quad& $\eta \gamma$ ($\omega$, $\phi$
dominated)& 0.70(02)\quad\\
$\eta \pi^+ \pi^-$& 1.34(05)\quad& $\eta \pi^+\pi^-\pi^0$ (no $\omega$)& 
0.71(08)\quad\\
$\eta 2\pi^+ 2\pi^-$& 0.08(01)\quad& $\eta \omega$& 0.30(02)\quad\\
$\eta \pi^+\pi^- 2\pi^0$& 0.12(02)\quad& $\omega (\rightarrow npp )2\pi$&
0.13(01)\quad\\
$\omega (\rightarrow \pi^0\gamma)\pi^0$& 0.88(02)\quad& $\omega 2\pi^+ 2\pi^-$&
0.01(00)\quad\\
$\omega (\rightarrow npp)3\pi$& 0.17(03)\quad& $\eta \phi$& 0.41(02)\quad\\
$\omega \eta \pi^0$& 0.24(05)\quad& $\phi \rightarrow unaccounted$& \, 0.04(04)
\quad\\
\hline
TOTAL:& 543.21(2.09)\quad&TOTAL:& 55.10(96)\quad\\
\hline
\end{tabular}
\caption{\label{tab1} {\it $G$-parity-unambiguous exclusive-mode contributions 
to $a_\mu^{\mathrm{LO,HVP}}$ for $\sqrt{s}\leq 1.937$~{\rm GeV} from KNT2019. 
Entries in units of $10^{-10}$. The notation ``npp'' is KNT2019's shorthand
for ``non-purely-pionic''.}}
\end{center}
\end{table}

\begin{boldmath}
\subsection{$G$-parity-unambiguous modes}
\end{boldmath}
The results of Table~I of KNT2019 for the $G$-parity-unambiguous 
exclusive-mode contributions to $a_\mu^{\mathrm{LO,HVP}}$ from the region 
between threshold and $\sqrt{s}=1.937$~GeV are tabulated in Table~\ref{tab1}.
From this table, the contribution of all $G$-parity-unambiguous exclusive 
modes to the sum of the strange-quark connected and disconnected 
contributions to $a_\mu^{\mathrm{LO,HVP}}$ is
\begin{equation}
\left[ 55.10(96)-{\frac{543.21(2.09)}{9}}\right]\times 10^{-10}\, =\,
-5.26(99)\times 10^{-10}\ .
\label{gparunambigsconnpdisc}
\end{equation}

\begin{boldmath}
\subsection{$K\bar{K}$-modes}
\end{boldmath}
The sum of the two $K\bar{K}$-mode contributions to 
$a_\mu^{\mathrm{LO,HVP}}$ in KNT2019, $(23.03(22)+13.04(19))\times 10^{-10}$, 
is sufficently large that the error produced by the ``maximally conservative'' 
$I=0/1$ separation assessment would be far too large to make 
the resulting determination of the disconnected contribution useful. 
The $K\bar{K}$ contributions are, however, expected to be 
dominated by those of the $I=0$ $\phi$ resonance, and the measured 
cross sections for both charge modes do, in keeping with this 
expectation, show very large peaks in the $\phi$ region. We need to 
turn this qualitative expectation into something more quantitative. 
This can be done using CVC and recent BaBar results for the 
unit-normalized differential $\tau\rightarrow K\bar{K}\nu_\tau$ 
decay distribution~\cite{BaBar:2018qry}. The latter, normalized to 
reproduce the corresponding branching fraction, gives an 
experimental determination of the charged $I=1$, vector-current 
spectral function~\cite{Tsai:1971vv} (the isospin partner of 
$\rho_{\mathrm{EM}}^{I=1}(s)$), and hence provides a determination
of the $I=1$ $e^+ e^-\rightarrow K\bar{K}$ contribution to $R(s)$, 
in the region $s< m_\tau^2$ kinematically accessible in $\tau$ 
decay.\footnote{This is up to (for our 
purposes) numerically negligible isospin-breaking corrections.} 
We have used the BaBar results to carry out this determination 
up to $s=2.7556$~GeV$^2$ (which corresponds to using all but the 
last BaBar bin, which is not used because of its large width and 
gigantic statistical error). This produces a contribution of
$0.764(9)(26)(18)\times 10^{-10}$ to $[a_\mu^{I=1}]_{K\bar{K}}$
from the region $s\leq 2.7556$~GeV$^2$, where the first error is
statistical, the second is systematic and the third is that induced 
by the uncertainty on the $\tau\rightarrow K^- K^0\nu_\tau$ branching 
fraction (taken from the HFLAV 2019 compilation~\cite{HFLAV:2019otj}),
which sets the overall normalization of the $\tau$-decay distribution. 
Above this point, we switch back to using $e^+e^-\rightarrow K\bar{K}$ 
cross-section data.{\footnote{Explicitly, we use the results
for the exclusive-mode $e^+e^-\rightarrow K^+ K^-$ and
$e^+e^-\rightarrow K_S K_L$ cross sections and covariances
employed by KNT2019, provided to us by the authors.}}
Integrating the KNT2019 2-mode, $I=0+1$ 
$e^+ e^-\rightarrow K\bar{K}$ cross-section sum from $s=2.7556$~GeV$^2$ 
to $(1.937~\mbox{GeV})^2=3.7520~\mbox{GeV}^2$, we find a ``maximally 
conservative error'' assessment of the $I=1$ contribution from this 
region of $0.089(89)\times 10^{-10}$.{\footnote{Additional
constraints, generated using experimental $K^+ K^-$ to $K_SK_L$ 
electroproduction cross-section ratios, in fact, allow this ``maximally 
conservative,'' $0.089\times 10^{-10}$, $I=0/1$ separation error to be 
reduced somewhat, to $0.065\times 10^{-10}$. Since, however, neither 
the improved nor the larger ``maximally conservative'' error is 
relevant on the scale of the uncertainty in our final results for 
$a_\mu^{\mathrm{sconn+disc}}$ and $a_\mu^{\mathrm{disc}}$, we do not 
discuss this improvement further, and employ the weaker ``maximally 
conservative'' error assessment in obtaining our final results below.}}
Combining this result with that from the region below 
$s= 2.7556$~GeV$^2$, we find a total $I=1$ $K\bar{K}$ contribution 
to $a_\mu^{\mathrm{LO,HVP}}$ from the region 
$s\leq (1.937~\mbox{GeV})^2=3.7520$~GeV$^2$ of
\begin{equation}
[a_\mu^{I=1}]_{K\bar{K}}=0.852(94)\times 10^{-10}\ .
\label{fullkkbaramuieq1contrib}\end{equation}
The central value in Eq.~(\ref{fullkkbaramuieq1contrib}) is dominated by 
the contribution from the region up to $2.7556$~GeV$^2$ determined by the 
BaBar $\tau$-decay data.

With KNT2019 giving a $\sqrt{s}\leq 1.937$~GeV $I=0+1$, two-mode
$K\bar{K}$ total of 
\begin{equation}
[a_\mu^{\mathrm{LO,HVP}}]_{K\bar{K}}=(23.03(22)+13.04(19))\times 10^{-10} 
= 36.07(29)\times 10^{-10}\, ,
\label{knt2019kkbartotalamu}\end{equation}
we find an $I=0$ contribution of $35.22(30)\times 10^{-10}$ and hence a 
$K\bar{K}$ strange-quark-connected plus disconnected contribution from the
region up to $\sqrt{s}=1.937$~GeV of
\begin{eqnarray}
[a_\mu^{\mathrm{sconn+disc}}]_{K\bar{K}}&&\equiv [a_\mu^{I=0}]_{K\bar{K}}
\, -\, {\frac{1}{9}}[a_\mu^{I=1}]_{K\bar{K}} \nonumber\\
&&= [a_\mu^{\mathrm{LO,HVP}}]_{K\bar{K}} 
-{\frac{10}{9}}[a_\mu^{I=1}]_{K\bar{K}}\,
 =\, 35.12(31)\times 10^{-10}\ .
\label{amukkbarsconnpdisc}
\end{eqnarray}
The error on this result, $0.31\times 10^{-10}$, is dramatically
reduced compared to the $\sim 20\times 10^{-10}$ uncertainty that would
result were the BaBar $\tau$-decay distribution results not available
and one were forced to rely on the ``maximally conservative'' assessment.

\vskip0.8cm
\begin{boldmath}
\subsection{$K\bar{K}\pi$-modes}
\end{boldmath}
The separation of the exclusive-mode $K\bar{K}\pi$ cross-sections
into $I=0$ and $I=1$ components was carried out already in 2007 by
BaBar~\cite{BaBar:2007ceh}, using a Dalitz plot analysis predicated on
the observed saturation of the cross sections by $KK^*$ contributions.
This result, combined with CVC, was actually already used, again long ago, 
by ALEPH to provide a data-based separation of $I=1$ vector and axial vector
contributions to the experimental $\tau\rightarrow K\bar{K}\pi\nu_\tau$
distribution~\cite{Davier:2008sk}. Here we simply convert the 
BaBar $I=1$ cross-section results to the corresponding $K\bar{K}\pi$
contributions to $R(s)$ and integrate these with the dispersive weight 
to determine the $I=1$ part of the $K\bar{K}\pi$ contribution to
$a_\mu^{\mathrm{LO,HVP}}$. The result is a contribution from the region 
up to $\sqrt{s}=1.937\ GeV$ of
\begin{equation}
[a_\mu^{I=1}]_{K\bar{K}\pi} = 0.741(36)(117) \times 10^{-10} 
= 0.74(12) \times 10^{-10}\, ,
\label{kkbarpiieq1amunlohvp}\end{equation}
where the first error is statistical and the second systematic.
With KNT2019 giving an $I=0+1$ total from this same region 
$[a_\mu^{\mathrm{LO,HVP}}]_{K\bar{K}\pi}=2.71(12) \times 10^{-10}$, we find 
a strange-quark-connected plus disconnected contribution of 
\begin{equation}
[a_\mu^{\mathrm{sconn+disc}}]_{K\bar{K}\pi} = 
[a_\mu^{\mathrm{LO,HVP}}]_{K\bar{K}\pi}
-{\frac{10}{9}}[a_\mu^{I=1}]_{K\bar{K}\pi}\, =\, 1.89(18) \times 10^{-10}\, .
\label{kkbarpisconnpdiscamulohvp}\end{equation}
For completeness, the corresponding $I=0$ contribution is 
$[a_\mu^{I=0}]_{K\bar{K}\pi}= 1.97(17)\times 10^{-10}$.

\vskip0.8cm
\subsection{The remaining $G$-parity-ambiguous modes}
The determination of the sum of the strange-quark-connected plus 
disconnected contributions from all remaining $G$-parity-ambiguous 
exclusive modes is detailed in the Appendix. Errors are, in all cases, 
obtained using the ``maximally conservative'' strategy. The result
for this sum, $1.05(98)\times 10^{-10}$ has a central value and 
uncertainty both dominated by the corresponding $K\bar{K}2\pi$ mode 
contribution.

\subsection{Perturbative contributions}\label{pcont}
Finally, we consider the pQCD contribution for the inclusive region 
(above $s=(1.937$~GeV$)^2=3.7520$~GeV$^2$).
To set notation, the pQCD expression for the EM current Adler function, 
$D(Q^2)\, =\, -Q^2 d\Pi_{\mathrm{EM}}(Q^2)/Q^2$, is~\cite{Baikov:2008jh}
\begin{eqnarray}
D(Q^2)&&={\frac{1}{6\pi^2}}\, \left[ 1+a(Q^2)+1.63982 a^2(Q^2)
+6.37101 a^3(Q^2)\right. \nonumber\\
&&\left. \quad +49.0757 a^4(Q^2)+d_5 a^5(Q^2)+\cdots \right]\ ,
\label{emadler}
\end{eqnarray}
where $a(Q^2)=\alpha_s(Q^2)/\pi$. The six-loop coefficient, $d_5$, is 
not yet known. In what follows we employ $d_5=283$~\cite{Beneke:2008ad}, 
which is in agreement with recent estimates of this 
coefficient~\cite{Boito:2018rwt,Caprini:2019kwp}.
With this notation, the corresponding pQCD expression for 
$\rho_{\mathrm{EM}}^{\mathrm{sconn+disc}}(s)$ is
\begin{eqnarray}
&&\left[ \rho_{\mathrm{EM}}^{\mathrm{sconn+disc}}(s)\right]_{\mathrm{pQCD}}=
{\frac{1}{6}}\, \left[ \rho_{\mathrm{EM}}(s)\right]_{\mathrm{pQCD}}\nonumber\\
&&\qquad\qquad =
{\frac{1}{36 \pi^2}}\, \left[ 1 + a(s) + 1.63982 a^2(s)
-10.2839 a^3(s)\right. \nonumber\\
&&\left. \qquad\qquad\qquad - 106.880 a^4(s) + (d_5-779.581)a^5(s)
+\cdots \right]\ .
\label{pqcdsconnpdisc}\end{eqnarray}
Integrating this numerically, using as input the $n_f=3$ result 
$\alpha_s(m_\tau^2)=0.3139(71)$, which follows from the 2020 PDG 
$n_f=5$ $M_Z$ scale value $0.1179(10)$, as an example, and keeping the 
6-loop term with the Beneke-Jamin $d_5=283$ estimate, gives a total 
inclusive-region $n_f=3$ contribution to $a_\mu^{\mathrm{sconn+disc}}$ of 
$6.281(5)\times 10^{-10}$. If we turn off all $O(\alpha_s^5)$ contributions, 
this becomes $6.291(6)\times 10^{-10}$. The quoted uncertainties are those 
associated with that on the PDG input for $\alpha_s$, and are so small we 
ignore them in what follows. The $\sim 0.01\times 10^{-10}$ difference 
between the fully known 5-loop- and estimated 6-loop-truncated results 
similarly suggests the uncertainty associated with truncating the pQCD 
expansion at five loops is negligible on the scale of other errors in 
our determinations of $a_\mu^{\mathrm{sconn+disc}}$ and 
$a_\mu^{\mathrm{disc}}$.

$D=2$ perturbative corrections in this region will be very small for the
following reasons. First, the light-quark loop contributions to the $I=0$ 
and $I=1$ $D=2$ perturbative series cancel in the $(I=0)$-$(1/9)(I=1)$ 
combination, and the MI contribution vanishes. This leaves a $D=2$ 
expression for the desired strange-quark-connected plus disconnected 
combination that is entirely that of the strange current $I=0$ contribution. 
The resulting $D=2$ contribution to the desired polarization combination 
is then easily read off from, for example, the vector current $D=2$ 
expression in Ref.~\cite{Chetyrkin:1993hi}, and is (neglecting $m_{u,d}$),
\begin{equation}
\left[\Pi_{\mathrm{EM}}^{\mathrm{sconn+disc}}(Q^2)\right]_{D=2}=\, -
{\frac{1}{6\pi^2}}{\frac{m^2_s(Q^2)}{Q^2}}\left[ 1+{\frac{8}{3}}
a(Q^2)+24.1415\, a^2(Q^2)+\cdots \right]\, .
\end{equation}
From this it is easily shown that the $D=2$ contribution to the perturbative 
representation of $\rho_{\mathrm{EM}}^{\mathrm{sconn+disc}}(s)$ is
\begin{equation}
\left[\rho_{\mathrm{EM}}^{\mathrm{sconn+disc}}(s)\right]_{D=2}=
{\frac{1}{6\pi^2}}{\frac{m^2_s(s)}{s}}\left[ 2a(s)+
{\frac{227}{12}}a^2(s)+\cdots \right]\, .
\end{equation}
The leading term in the $D=2$ series is thus a factor of
$12m_s^2(s)a(s)/s$ times the leading term in the $D=0$ series.
This represents a factor of $\sim 1/400$ suppression of the leading 
$D=2$ relative to leading $D=0$ contribution for $s\sim 4$~GeV$^2$. 
This rough estimate for the size of the integrated inclusive-region 
$D=2$ contribution relative to the corresponding integrated $D=0$ 
contribution is born out by the result of integrating the 
three-loop-truncated $D=2$ series, with integration from 
$s=(1.937\ {\mathrm{GeV}})^2$ to infinity producing a $D=2$ contribution 
to $a_\mu^{\mathrm{sconn+disc}}$ of $0.012\times 10^{-10}$, to be 
compared to the result $6.29\times 10^{-10}$ for the corresponding 
five-loop-truncated $D=0$ contribution.

Still to be carried out is an estimate of the impact of possible small 
residual exponentially damped quark-hadron duality-violating (DV) 
contributions to $\rho_{\mathrm{EM}}^{\mathrm{sconn+disc}}(s)$. We 
expect these to also be small, though not as small as the tiny $D=2$ 
corrections. The question of the size of possible DV-induced 
uncertainties can be explored by updating recent FESR fits to 
electroproduction cross-section data and $I=1$ $\tau$-decay data, 
in which the parameters of a large-$N_c$+Regge-motivated 
{\it ansatz}~\cite{Boito:2017cnp} for the $s$-dependence of the DV
contribution to the vector $I=1$ and $I=0$ spectral functions are 
obtained as part of the fits. We return to this point in 
Sec.~\ref{discussion} below.

\vskip1cm
Combining the $G$-parity-unambiguous mode contributions
given in Eq.~(\ref{gparunambigsconnpdisc}) with the $K\bar K$ and 
$K\bar K\pi$ contributions given in Eqs.~(\ref{amukkbarsconnpdisc}) 
and~(\ref{kkbarpisconnpdiscamulohvp}), the remaining $G$-parity-ambiguous 
mode contributions (dominated by $K\bar K 2\pi$) detailed in the
Appendix, and the inclusive region contribution, estimated using
pQCD, detailed in Sec.~\ref{pcont}, we obtain the following result 
for the sum of strange-quark-connected and disconnected contributions 
to $a_\mu^{\mathrm{LO,HVP}}$:
\begin{eqnarray}
a_\mu^{\mathrm{sconn+disc}}&&=\left[ 55.10(96)-{\frac{543.21(2.09)}{9}}
+35.12(31)+1.89(18)+1.05(98)+6.28\right]\times 10^{-10}\nonumber\\
&&=39.08(1.44)\times 10^{-10}\ .
\label{sconnpdiscnoibcorr}\end{eqnarray}

Subtracting from this the $g-2$ Theory Initiative white paper average,
$a_\mu^{\mathrm{sconn}}=53.2(3)\times 10^{-10}$~\cite{Aoyama:2020ynm}, 
for the strange-quark-connected contribution, obtained by averaging the 
results of Refs.~\cite{Chakraborty:2014mwa,RBCUKQCD:2016clu,Budapest-Marseille-Wuppertal:2017okr,RBC:2018dos,Shintani:2019wai,Gerardin:2019rua,Giusti:2018mdh,Borsanyi:2020mff},
we find as our initial estimate for the disconnected contribution 
\begin{equation}
a_\mu^{\mathrm{disc}}\, =\, -14.1(1.5)\times 10^{-10}\ .
\label{amudiscnoibcorr}
\end{equation}
Expanding the whitepaper average to include the new BMW result, 
$a_\mu^{\mathrm{sconn}}=53.393(89)(68)\times
10^{-10}$~\cite{Borsanyi:2020mff}, shifts this estimate only slightly, to 
\begin{equation}
a_\mu^{\mathrm{disc}}\, =\, -14.3(1.4)\times 10^{-10}\ .
\label{amudiscnoibcorrinclbmw20sconn}
\end{equation}

The disconnected result, Eq.~(\ref{amudiscnoibcorrinclbmw20sconn}),
is, of course, obtained without having to carry out the numerically 
intensive determination of disconnected contributions on the lattice. 
It is, however, not yet directly comparable to current lattice 
determinations, since the latter are defined in the isospin limit. 
The exclusive-mode contributions used in obtaining the estimate of
Eq.~(\ref{amudiscnoibcorr}), in contrast, are physical ones and will 
include small IB contributions. In the next section 
we explore the size of possible IB corrections to the inital estimates 
for $a_\mu^{\mathrm{sconn+disc}}$ and $a_\mu^{\mathrm{disc}}$ given in 
Eqs.~(\ref{sconnpdiscnoibcorr}) and (\ref{amudiscnoibcorrinclbmw20sconn}).

\section{Estimates for isospin-breaking corrections}\label{ibcorrns}
To make contact with lattice determinations of $a_\mu^{\mathrm{sconn+disc}}$ 
and $a_\mu^{\mathrm{disc}}$, it is necessary to estimate and subtract
IB contributions to the results of Eqs.~(\ref{sconnpdiscnoibcorr}) 
and (\ref{amudiscnoibcorrinclbmw20sconn}). We consider strong and EM IB 
corrections separately. The recent BMW lattice paper~\cite{Borsanyi:2020mff} 
is the first to provide determinations of all EM contributions, and, to 
take advantage of those results in the discussion below, we work with 
the SIB contribution defined in the same scheme for separating strong 
and EM IB as used by BMW (defined such that all EM effects in the purely 
connected neutral pseudoscalar masses are absorbed into the definitions 
of the quark masses, and numerically very similar to the GRS 
scheme~\cite{Gasser:2003hk}).{\footnote{A clear discussion of the 
SIB/EM separation issue can be found in Sections 3.1.1 and 3.1.2 of 
the 2019 FLAG report~\cite{FlavourLatticeAveragingGroup:2019iem}.}}

To determine the IB corrections to the initial estimates for 
$a_\mu^{\mathrm{sconn+disc}}$ and $a_\mu^{\mathrm{disc}}$ obtained above 
and make contact with isospin-limit lattice results, one needs to identify 
and subtract the IB contributions present in the experimental versions of 
the nominal $I=0$ and $I=1$ contributions, $a_\mu^{I=0}$ and $a_\mu^{I=1}$, 
determined above. These are of two types: those belonging to the 
mixed-isospin ($a_\mu^{\mathrm{MI}}$) ``contaminations'' of the 
various physical exclusive-mode contributions, and those present in the 
physical $a_\mu^{I=0}$ and $a_\mu^{I=1}$ contributions themselves. An 
inclusive determination of the IB contributions present in the physical 
versions of $a_\mu^{I=0}$ and $a_\mu^{I=1}$ is sufficient to perform the 
latter correction. Correcting for the mixed-isospin contaminations, 
however, requires identifying the mixed-isospin components of the 
various exclusive-mode $a_\mu^{\mathrm{LO,HVP}}$ contributions. 

A rough (likely conservative) estimate for the scale of the IB-induced 
uncertainty can be obtained by assuming an $O(1\%)$ scale for IB 
corrections to each of the non-pQCD terms in Eq.~(\ref{sconnpdiscnoibcorr}),
and adding these uncertainties linearly (corresponding to the presumably 
conservative assumption that all enter the combination in 
Eq.~(\ref{sconnpdiscnoibcorr}) with the same sign). The result is a 
first-pass estimate of $\sim 1.5\times 10^{-10}$ for the IB-induced 
uncertainty on the result for $a_\mu^{I=0}-{\frac{1}{9}} a_\mu^{I=1}$. 
We thus do not expect IB corrections to dramatically shift the results 
for $a_\mu^{\mathrm{sconn+disc}}$ or $a_\mu^{\mathrm{disc}}$ obtained 
above. An improved estimate of the IB corrections can be obtained as 
discussed below.

To first order in IB, SIB appears only in the mixed-isospin part of the EM 
polarization/EM spectral function. Thus, again to first order in IB, the IB 
contributions present in the physical versions of $a_\mu^{I=0}$ and 
$a_\mu^{I=1}$ are purely EM in nature. We need these only in an inclusive 
(sum over exclusive-modes) form, and will take advantage of details 
of the recent BMW lattice assessment of EM contributions to 
$a_\mu^{\mathrm{LO,HVP}}$~\cite{Borsanyi:2020mff} to both show these
must be small and obtain actual estimates of their size, confirming 
this expectation, as described in more detail below.
We will then turn to the issue of the mixed-isospin ``contaminations''
present in the physical exclusive-mode contributions.

\subsection{EM IB contributions}

\begin{figure}[t]
\begin{center}
\includegraphics[width=.33\textwidth,angle=270]
{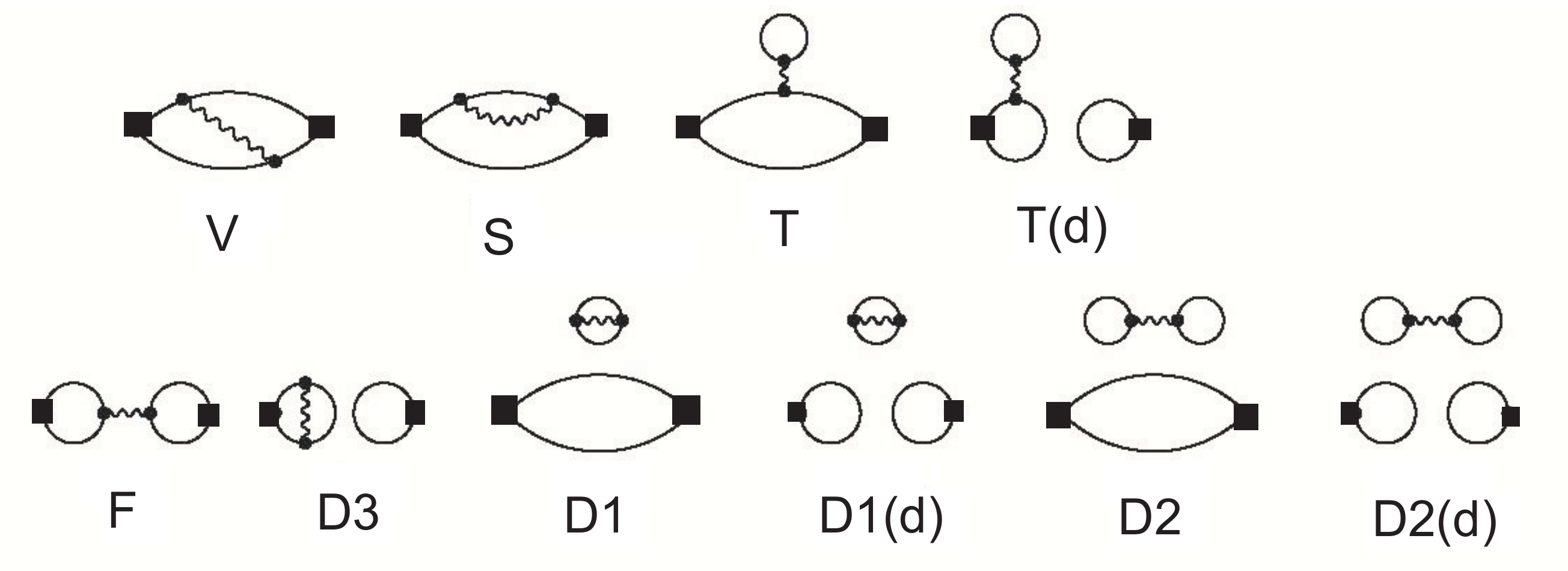}
\caption{\label{emibgraphs} {\it Valence-valence ($vv$), valence-sea ($vs$)
and sea-sea ($ss$) connected ($c$) and disconnected ($d$) graphs
contributing to the the full, $I=1$, $I=0$ and $MI$ EM contributions
to $a_\mu^{\mathrm{LO,HVP}}$, following the RBC/UKQCD labelling scheme
\cite{Gulpers:2018mim}. For a more detailed description, see the main
text.}}
\end{center}
\end{figure}

The diagrams producing $O(\alpha_{\mathrm{EM}})$ contributions to 
$\hat{\Pi}_{\mathrm{EM}}$, $\hat{\Pi}_{\mathrm{EM}}^{I=1}$, 
$\hat{\Pi}_{\mathrm{EM}}^{I=0}$ and $\hat{\Pi}_{\mathrm{EM}}^{\mathrm{MI}}$, 
and hence $O(\alpha^2_{\mathrm{EM}})$ contributions to
$a_\mu^{\mathrm{LO,HVP}}$, $a_\mu^{I=1}$, $a_\mu^{I=0}$ and 
$a_\mu^{\mathrm{MI}}$, are shown schematically in Fig.~\ref{emibgraphs}. 
The labelling follows the conventions of RBC/UKQCD~\cite{Gulpers:2018mim}. 
The black squares denote external-current vertices. Gluon lines are not 
shown explicitly, but are to be understood as connecting all quark lines 
in each diagram. The external-current couplings to a flavor $k=u,\, d,\, s$ 
quark loop are $c^a_k$ ($k=u,d,s$) if the external current is one of 
$J_\mu^{\mathrm{EM},a}$ ($a=3$ or $8$), and 
$c_k^{\mathrm{EM}}=Q_k=c^3_k+c^8_k$, with $Q_k$ the quark charge in 
units of $e$, if the external current is $J_\mu^{\mathrm{EM}}$. Since 
$\sum_{k=u,d,s}c_k^3 =\sum_{k=u,d,s}c^8_k=\sum_{k=u,d,s}c_k^{\mathrm{EM}}=0$, 
diagrams containing a quark loop with only one vertex, whether of 
external-current or internal-photon type, vanish in the flavor-$SU(3)$ 
($SU(3)_F$) limit. Diagrams $V$, $S$, $F$ and $D1$ have no such suppression 
and survive in the $SU(3)_F$ limit. Diagrams $T$ and $D3$ have a single 
such suppression, diagrams $T(d)$, $D1(d)$ and $D2$ a double suppression, 
and diagram $D2(d)$ a four-fold suppression. The contributions fall into 
valence-valence ($vv$), valence-sea ($vs$) and sea-sea ($ss$) subsets, 
the double label specifying whether the internal-photon line connects 
two valence-quark lines, one valence- and one sea-quark line, or two 
sea-quark lines. These subsets can be further broken down into connected 
($c$) and disconnected ($d$) parts, characterized by whether the two 
external vertices lie on the same or different quark loops. The 
$(vv,c)$, $(vv,d)$, $(vs,c)$, $(vs,d)$, $(ss,c)$ and $(ss,d)$ 
contributions are the sums of contributions from diagrams $V+S$, $F+D3$, 
$T$, $T(d)$, $D1+D2$ and $D1(d)+D2(d)$, respectively. Both $(vv,c)$ 
diagrams are non-zero in the $SU(3)_F$ limit. The $(vv,d)$ contribution 
is expected to be dominated by the unsuppressed diagram-$F$ contribution 
and the $(ss,c)$ contribution by the unsuppressed diagram-$D1$ 
contribution.{\footnote{Diagrams $F$ and $D1$ are $1/N_c$ suppressed,
and the BMW results, indeed, show a suppression of the contributions 
from these diagrams relative to the $1/N_c$-unsuppressed $V+S$ 
contribution sum.}} All other contributions, including the full 
$(vs,c)$, $(vs,d)$ and $(ss,d)$ combinations, vanish in the $SU(3)_F$ limit. 

BMW~\cite{Borsanyi:2020mff} has provided the following lattice results 
for the $vv$, $vs$, $ss$ connected and disconnected contributions to the 
full EM-current result, $a_\mu^{\mathrm{LO,HVP}}$:
\begin{eqnarray}
&&\left[ a_\mu^{\mathrm{LO,HVP}}\right]_{(vv,c)}\, =\, 
-1.23(40)(31)\times 10^{-10}
\qquad (V+S)\ ,\\
&&\left[ a_\mu^{\mathrm{LO,HVP}}\right]_{(vv,d)}\, =\, 
-0.55(15)(10)\times 10^{-10}
\qquad (F+D3)\ ,\nonumber\\
&&\left[ a_\mu^{\mathrm{LO,HVP}}\right]_{(vs,c)}\, =\, 
-0.0093(86)(95)\times 10^{-10}
\quad (T)\ ,\nonumber\\
&&\left[ a_\mu^{\mathrm{LO,HVP}}\right]_{(vs,d)}\, =\, 
\ \ 0.011(24)(14)\times 10^{-10}
\quad\ \  (T(d))\ ,\nonumber\\
&&\left[ a_\mu^{\mathrm{LO,HVP}}\right]_{(ss,c)}\, =\, 
\ \ 0.37(21)(24)\times 10^{-10}
\qquad\, (D1+D2)\ ,\nonumber\\
&&\left[ a_\mu^{\mathrm{LO,HVP}}\right]_{(ss,d)}\, =\, 
-0.040(33)(21)\times 10^{-10}
\quad\ \,  (D1(d)+D2(d))\nonumber\ .
\end{eqnarray}
The $SU(3)_F$-suppressed $(vs,c)$, $(vs,d)$ and $(ss,d)$ contributions
are a factor of $\sim~10$ or more smaller than their $SU(3)_F$-unsuppressed
counterparts, $(vv,c)$, $(vv,d)$ and $(ss,c)$, suggesting a strong suppression
in differences of light- and strange-quark-loop contributions, further 
supporting the expectation that diagram $F$ will dominate the $(vv,d)$ 
contribution and diagram $D1$ the $(ss,c)$ contribution.

Based on this observation, we will, in what follows, neglect contributions 
which vanish in the $SU(3)_F$ limit, and discuss the extent to which the 
BMW full-EM-current results can be broken down into their $I=1$, $I=0$ and 
mixed-isospin $(MI$) components. Such a breakdown will allow us to estimate 
the desired EM contributions to $a_\mu^{\mathrm{MI}}$ and 
$a_\mu^{\mathrm{sconn+disc}}=a_\mu^{I=0}-{\frac{1}{9}}a_\mu^{I=1}$.

It is useful to first introduce some notation for the loop contributions 
to the various subtracted polarizations from the unsuppressed diagrams of 
Fig.~\ref{emibgraphs}. It is convenient to explicitly factor out (i) both the 
external-current couplings and the internal-photon couplings (the latter 
in units of $e$) for diagrams $V$, $S$ and $F$ and (ii) the external-current 
couplings for diagram $D1$. In a given lattice configuration, the loop 
contributions are determined by the quark propagators, which are in turn 
fixed by the gluon field configuration. The corresponding contributions to 
the subtracted polarizations are obtained by averaging over the ensemble. We 
define the ``loop factor'' for a given contribution as the full contribution 
divided (i) by the product of the relevant explicit external-current and 
internal-photon couplings for diagrams $V$, $S$ and $F$ and (ii) by the 
product of the relevant explict external-current couplings for diagram 
$D1$.{\footnote{The factors of $e^2$ coming from the two internal-photon 
vertices and the relevant internal-photon propagators are thus, in all 
cases, absorbed into the definitions of the corresponding loop factors.}} 
The following notation is employed for the loop factors of the 
$SU(3)_F$-unsuppressed contributions from the connected diagrams 
$\kappa = V,\, S$ and $D1$ and disconnected diagram $F$: 
\begin{itemize}
\item $L_j^{(\kappa )}$, $j=u,d,s$, denotes the flavor-$j$
quark-loop factor for connected diagram $\kappa =V,\, S$. With this notation, 
the diagram-$V$ contribution to $\hat{\Pi}_{\mathrm{EM}}$, for example, is
\begin{equation}
\sum_{k=u,d,s}Q_k^4 L_k^{(V)}\ ,
\end{equation}
where two factors of $Q_k$ come from the external-current vertices and the
remaining two from the internal-photon vertices. Similarly, the diagram-$V$ 
contribution to $\hat{\Pi}^{I=1}$ is
\begin{equation}
\sum_{k=u,d,s}\left(c_k^{\mathrm{\mathrm{EM}},3}\right)^2 Q_k^2\, L_k^{(V)}\ ,
\end{equation}
that to $\hat{\Pi}^{I=0}$,
\begin{equation}
\sum_{k=u,d,s}\left(c_k^{\mathrm{\mathrm{EM}},8}\right)^2 Q_k^2\, L_k^{(V)}\ ,
\end{equation}
and that to $\hat{\Pi}^{\mathrm{MI}}$
\begin{equation}
2\, \sum_{k=u,d,s}c_k^{\mathrm{\mathrm{EM}},3}c_k^{\mathrm{\mathrm{EM}},8} 
Q_k^2\, L_k^{(V)}\ ,
\end{equation}
where the factor of $2$ results from the presence of both $ab=38$ and $83$ 
contributions.
\item $LL_{k,m}^{(F)}$ denotes the loop factor for the two-disconnected-loop 
diagram-$F$ contribution with a flavor-$k$ quark loop attached to the left 
external vertex and a flavor-$m$ quark loop attached to the right external 
vertex. This represents the ensemble average over the product of the two loops,
and does not factorize into a product of the ensemble averages of the 
individual loops. With this definition, the diagram-$F$ contribution
to $\hat{\Pi}_{\mathrm{EM}}$ is
\begin{equation}
\sum_{k,m=u,d,s}Q_k^2Q_m^2\, LL_{k,m}^{(F)}\ ,
\end{equation}
that to $\hat{\Pi}_{\mathrm{EM}}^{I=1}$,
\begin{equation}
\sum_{k,m=u,d,s}c_k^{\mathrm{EM},3} c_m^{\mathrm{EM},3} Q_k Q_m 
LL_{k,m}^{(F)}\ ,
\end{equation}
that to $\hat{\Pi}_{\mathrm{EM}}^{I=0}$,
\begin{equation}
\sum_{k,m=u,d,s}c_k^{\mathrm{EM},8} c_m^{\mathrm{EM},8} Q_k Q_m 
LL_{k,m}^{(F)}\ ,
\end{equation}
and that to $\hat{\Pi}_{\mathrm{EM}}^{\mathrm{MI}}$,
\begin{equation}
2\, \sum_{k,m=u,d,s}c_k^{\mathrm{EM},3} c_m^{\mathrm{EM},8} Q_k Q_m 
LL_{k,m}^{(F)}\ .
\end{equation}
\item $LL_{k}^{(D1)}$ denotes the loop factor for the two-disconnected-loop 
diagram-$D1$ contribution with a flavor-$k$ quark loop attached to both 
external vertices, summed over all three flavors in the disconnected EM 
vacuum bubble. This again represents the ensemble average over the product 
of the two loops, and does not factorize into a product of the ensemble 
averages of the individual loops. With this definition, the diagram-$D1$ 
contribution to $\hat{\Pi}_{\mathrm{EM}}$ is
\begin{equation}
\sum_{k=u,d,s}Q_k^2\, LL_{k}^{(D1)}\ ,
\end{equation}
that to $\hat{\Pi}_{\mathrm{EM}}^{I=1}$,
\begin{equation}
\sum_{k=u,d,s}\left( c_k^{\mathrm{EM},3}\right)^2 LL_{k}^{(D1)}\ ,
\end{equation}
that to $\hat{\Pi}_{\mathrm{EM}}^{I=0}$,
\begin{equation}
\sum_{k=u,d,s}\left( c_k^{\mathrm{EM},8}\right)^2 LL_{k}^{(D1)}\ ,
\end{equation}
and that to $\hat{\Pi}_{\mathrm{EM}}^{\mathrm{MI}}$,
\begin{equation}
2\, \sum_{k=u,d,s}c_k^{\mathrm{EM},3} c_k^{\mathrm{EM},8} LL_{k}^{(D1)}\ .
\end{equation}
\end{itemize}
The $\hat{\Pi}_{\mathrm{EM}}$ contributions are, of course, in all cases 
the sums of the corresponding $\hat{\Pi}_{\mathrm{EM}}^{I=1}$, 
$\hat{\Pi}_{\mathrm{EM}}^{I=0}$ and $\hat{\Pi}_{\mathrm{EM}}^{\mathrm{MI}}$ 
contributions.

The associated contributions to $a_\mu^{\mathrm{LO,HVP}}$, $a_\mu^{I=1}$,
$a_\mu^{I=0}$ and $a_\mu^{\mathrm{MI}}$ are obtained by integrating the 
relevant subtracted polarization with respect to $Q^2$, using the 
Euclidean-$Q^2$ weighting given in Eq.~(\ref{amueuclhvplatt}) above. 
Expressions for these contributions can be obtained from the corresponding 
expressions for the contributions to the subtracted polarizations by 
replacing the loop factors, $L_k^{(\kappa )}$ ($\kappa =V,\, S$), 
$LL_{k,m}^{(F)}$ and $LL_{k}^{(D1)}$ with the correspondingly weighted 
integrals of these loop factors, which we denote by $\bar{L}_k^{(\kappa )}$ 
($\kappa =V,\, S$), $\overline{LL}_{k,m}^{(F)}$ and 
$\overline{LL}_{k}^{(D1)}$ in what follows.

With this notation, the breakdowns for the $SU(3)_F$-unsuppressed 
contributions of diagrams $V$, $S$, $F$ and $D1$ proceed as follows, where 
we take into account that the resulting decomposition is to be applied to 
the BMW results for the full $a_\mu^{\mathrm{LO,HVP}}$ $(vv,c)$, $(vv,d)$ 
and $(ss,c)$ contributions and that those results were obtained as 
corrections evaluated using isospin-symmetric configurations. Since 
all relevant flavor-dependent coupling factors were explicitly factored 
out in defining the loop factors for the subtracted polarizations above, 
one has $L_u^{(\kappa )}=L_d^{(\kappa )}\equiv L_\ell^{(\kappa )}$
for the diagram $\kappa = V,\, S$ loop factors, $LL_{u}^{(D1)}=LL_{d}^{(D1)}$
for the diagram $D1$ loop factors, 
and $LL_{u,u}^{(F)}=LL_{u,d}^{(F)} =LL_{d,u}^{(F)}=LL_{u,u}^{(F)}
\equiv LL_{\ell ,\ell }^{(F)}$ and $LL_{u,s}^{(F)}=LL_{d,s}^{(F)}
=LL_{s,u}^{(F)}=LL_{s,d}^{(F)}\equiv LL_{\ell ,s}^{(F)}$ 
for the diagram $F$ loop factors, with similar relations for the integrated 
loop factors $\bar{L}_k^{(\kappa )}$, $\overline{LL}_{k}^{(D1)}$ and
$\overline{LL}_{k,m}^{(F)}$.

Since diagrams $V$ and $S$ share the same products of external-current 
and internal-photon coupling factors, it is also convenient to define 
the combined $V+S\equiv (vv,c)$ loop factors,
\begin{equation}
\bar{L}_k^{(vv,c)}\equiv \bar{L}_k^{(V)}+\bar{L}_k^{(S)}\ .
\end{equation}
We also introduce the $SU(3)_F$-breaking ratios
\begin{eqnarray}
&&x_{(vv,c)} \equiv \bar{L}_s^{(vv,c)}/\bar{L}_\ell^{(vv,c)}\ ,\\
&&x_{D1} \equiv \overline{LL}_{s}^{(D1)}/\overline{LL}_{\ell}^{(D1)}
\ ,\nonumber\\
&&x_F \equiv \overline{LL}_{\ell ,s}^{(F)}/\overline{LL}_{\ell ,\ell }^{(F)}
\ ,\nonumber\\
&&y_F \equiv \overline{LL}_{s,s}^{(F)}/\overline{LL}_{\ell ,\ell }^{(F)}\ ,
\nonumber
\end{eqnarray}
for use in the expressions below. These ratios are all, of course, equal to
$1$ in the $SU(3)_F$ limit.{\footnote{Note that $SU(3)$ breaking for these 
quantities can be larger than naively expected since the weight in the 
dispersive representation emphasizes strongly the low-$s$ part of the 
spectrum. $SU(3)$ breaking in the subtracted polarizations (related to the 
spectral functions by the usual subtracted dispersion relation) can then be 
further enhanced by ``kinematic'' effects associated with the shift to higher 
$s$ in the spectrum (and hence reduced dispersive weight) of states which 
couple to the strange current compared to those which couple to the 
non-strange current. An example is provided by the relation between the 
strange and light ($ud$) connected components of the isospin-limit 
contributions to $a_\mu^{\mathrm{LO,HVP}}$. Using $C$ to denote the 
corresponding connected diagram (diagram $D1$ without the EM vacuum 
bubble), one has $\left[ a_\mu^{\mathrm{LO,HVP}}\right]_C= 
\left[ 5\bar{L}_\ell^{(C)}+\bar{L}_s^{(C)}\right] /9$. In the $SU(3)_F$ 
limit, where $x_C\equiv \bar{L}_s^{(C)}/\bar{L}_\ell^{(C)} =1$, the 
light-connected contribution is thus a factor of $5$ larger than the 
strange-quark-connected contribution, while for physical $m_s$ and $m_\ell$ 
this ratio is $\sim 12.2$~\cite{Aoyama:2020ynm}, corresponding to 
$x_C\simeq 0.41$.}}

With this notation established, one finds the following bounds:
\begin{itemize}
\item {\it For the $(vv,c)$ (diagrams $V+S$) sum}: 
\begin{eqnarray}
&&\left[ a_\mu^{\mathrm{LO,HVP}}\right]_{(vv,c)} = 
\left[ 17 \bar{L}_\ell^{(vv,c)}+\bar{L}_s^{(vv,c)}\right] /81\ ,\\
&&\left[ a_\mu^{I=1}\right]_{(vv,c)} = 5\bar{L}_\ell^{(vv,c)}/36\ ,\nonumber\\
&&\left[ a_\mu^{I=0}\right]_{(vv,c)} = \left[ 5\bar{L}_\ell^{(vv,c)}
+4\bar{L}_s^{(vv,c)}\right] /324
\ ,\nonumber\\
&&\left[ a_\mu^{\mathrm{MI}}\right]_{(vv,c)} = \left[ 
\bar{L}_\ell^{(vv,c)}\right] /18\ .
\nonumber
\label{amudiagvpscombos}\end{eqnarray}
The combinations of interest, 
$\left[ a_\mu^{\mathrm{sconn+disc}}\right]_{(vv,c)}$ 
and $\left[ a_\mu^{\mathrm{MI}}\right]_{(vv,c)}$, thus represent fractions 
\begin{eqnarray}
&&\left[ a_\mu^{\mathrm{sconn+disc}}\right]_{(vv,c)}/\left[ 
a_\mu^{\mathrm{LO,HVP}}\right]_{(vv,c)}
={\frac{1}{17}}\, {\frac{x_{(vv,c)}}{1+(x_{(vv,c)}/17)}}\ ,\\
&&\left[ a_\mu^{\mathrm{MI}}\right]_{(vv,c)}/
\left[ a_\mu^{\mathrm{LO,HVP}}\right]_{(vv,c)}
={\frac{9}{34}}\, {\frac{1}{1+(x_{(vv,c)}/17)}}\ ,\nonumber
\label{vvcfracs}\end{eqnarray}
of the BMW $(vv,c)$ total $-1.23(40)(31)\times 10^{-10}$. The cancellation
of the terms proportional to $\bar{L}_\ell^{(vv,c)}$ is a generic feature
of the $(vv,c)$ contribution 
$\left[ a_\mu^{\mathrm{sconn+disc}}\right]_{(vv,c)}$,
and produces a sizeable numerical suppression of this contribution relative 
to the full $(vv,c)$ sum. If we consider the (presumably conservative) range 
$0\leq x_{(vv,c)}\leq 1$, we see that 
$\left[ a_\mu^{\mathrm{sconn+disc}}\right]_{(vv,c)}$ should lie somewhere 
between $0$ and $1/18$ times the full BMW $(vv,c)$ result, {\it i.e.}, 
in the range 
\begin{equation}
-0.068(22)(17)\times 10^{-10}\leq \left[ 
a_\mu^{\mathrm{sconn+disc}}\right]_{(vv,c)}\leq 0\ ,
\label{vvcamusconnpdisc}\end{equation}
and hence produce an essentially negligible correction to the 
strange-quark-connected plus disconnected and disconnected results of
Eqs.~(\ref{sconnpdiscnoibcorr}) and (\ref{amudiscnoibcorrinclbmw20sconn}) 
above. For the same range of $x_{(vv,c)}$, the mixed-isospin $(vv,c)$ 
contribution lies in the range between $9/34$ and $1/4$ times the full 
$(vv,c)$ result, {\it i.e.}, in the rather precisely determined range
\begin{equation}
-0.326(106)(82)\times 10^{-10}\leq \left[ a_\mu^{\mathrm{MI}}\right]_{(vv,c)}
\leq -0.308(100)(78)\times 10^{-10}\ .
\label{vvcamumi}\end{equation}

\item {\it For the $(vv,d)$ contribution}: 
Neglecting the $SU(3)_F$-suppressed diagram-$D3$ contribution,
\begin{eqnarray}
&&\left[ a_\mu^{\mathrm{LO,HVP}}\right]_{(vv,d)} = 
\left[ 25 \overline{LL}_{\ell ,\ell}^{(F)}+10\overline{LL}_{\ell ,s}^{(F)}
+\overline{LL}_{s,s}^{(F)}\right] /81\ ,\\
&&\left[ a_\mu^{I=1}\right]_{(vv,d)} = \overline{LL}_{\ell ,\ell}^{(F)}/4
\ ,\nonumber\\
&&\left[ a_\mu^{I=0}\right]_{(vv,d)} = \left[ 
\overline{LL}_{\ell ,\ell}^{(F)}+4\overline{LL}_{\ell ,s}^{(F)}+
4\overline{LL}_{s,s}\right]/324\ , \nonumber\\
&&\left[ a_\mu^{\mathrm{MI}}\right]_{(vv,d)} = \left[ 
\overline{LL}_{\ell ,\ell}^{(F)}
+2\overline{LL}_{\ell ,s}^{(F)}\right]/18\ .\nonumber
\label{amudiagd3combos}\end{eqnarray}
The combinations of interest, 
$\left[ a_\mu^{\mathrm{sconn+disc}}\right]_{(vv,d)}$ 
and $\left[ a_\mu^{\mathrm{MI}}\right]_{(vv,d)}$, then represent fractions 
\begin{eqnarray}
&&\left[ a_\mu^{\mathrm{sconn+disc}}\right]_{(vv,d)}/\left[ 
a_\mu^{\mathrm{LO,HVP}}\right]_{(vv,d)}
=\, -\left[ {\frac{2-x_F-y_F}{25+10x_F+y_F}}\right]\ ,\\
&&\left[ a_\mu^{\mathrm{MI}}\right]_{(vv,d)}/
\left[ a_\mu^{\mathrm{LO,HVP}}\right]_{(vv,d)}
={\frac{9}{2}}\, \left[ {\frac{1+2x_F}{25+10x_F+y_F}}\right]\ ,\nonumber
\label{vvdfracs}\end{eqnarray}
of the BMW $(vv,d)$ total, $-0.55(15)(10)\times 10^{-10}$. There is a 
strong cancellation (exact in the $SU(3)_F$ limit) for the 
strange-quark-connected plus disconnected combination. Once more assuming 
the joint range $0\leq x_F\leq 1$, $0\leq y_F\leq 1$ to represent a 
conservative choice, $\left[ a_\mu^{\mathrm{sconn+disc}}\right]_{(vv,d)}$ 
is expected to lie somewhere between $-2/25$ and $0$ times the full BMW 
$(vv,d)$ result, {\it i.e.}, in the range 
\begin{equation}
0\leq \left[ a_\mu^{\mathrm{sconn+disc}}\right]_{(vv,d)} 
\leq 0.044(12)(8)\times 10^{-10}\ ,
\label{vvdamusconnpdisc}\end{equation}
and hence to again produce an essentially negligible correction to the 
strange-quark-connected plus disconnected and disconnected results,
Eqs.~(\ref{sconnpdiscnoibcorr}) and (\ref{amudiscnoibcorrinclbmw20sconn}),
above. For the same joint $x_F$, $y_F$ range, the mixed-isospin $(vv,d)$ 
contribution lies between $9/52$ and $27/70$ times the full $(vv,d)$ result, 
{\it i.e.}, in the range
\begin{equation}
-0.212(58)(39)\times 10^{-10}\leq \left[ a_\mu^{\mathrm{MI}}\right]_{(vv,d)}
\leq -0.095(26)(17)\times 10^{-10}\ .
\label{vvdamumi}\end{equation}

\item {\it For the $(ss,c)$ contribution}: 
Neglecting the doubly $SU(3)_F$-suppressed diagram-$D2$ contribution,
\begin{eqnarray}
&&\left[ a_\mu^{LO,HVP}\right]_{(ss,c)} = 
\left[ 5 \overline{LL}_\ell^{(D1)}+\overline{LL}_s^{(D1)}\right] /9\ ,\\
&&\left[ a_\mu^{I=1}\right]_{(ss,c)} = \overline{LL}_\ell^{(D1)}/2\ ,
\nonumber\\
&&\left[ a_\mu^{I=0}\right]_{(ss,c)} = \left[ \overline{LL}_\ell^{(D1)}
+2\overline{LL}_s^{(D1)}\right] /18\ ,
\nonumber\\
&&\left[ a_\mu^{\mathrm{MI}}\right]_{(ss,c)} = 0\ .\nonumber
\label{amudiagfcombos}\end{eqnarray}
The combinations of interest, 
$\left[ a_\mu^{\mathrm{sconn+disc}}\right]_{(ss,c)}$ 
and $\left[ a_\mu^{\mathrm{MI}}\right]_{(ss,c)}$, thus represent fractions 
\begin{eqnarray}
&&\left[ a_\mu^{\mathrm{sconn+disc}}\right]_{(ss,c)}/\left[ 
a_\mu^{\mathrm{LO,HVP}}\right]_{(ss,c)}
={\frac{1}{5}}\, {\frac{x_{D1}}{1+(x_{D1}/5)}}\ ,\\
&&\left[ a_\mu^{\mathrm{MI}}\right]_{(ss,c)}/
\left[ a_\mu^{\mathrm{LO,HVP}}\right]_{(ss,c)}
=0\ ,\nonumber
\label{sscfracs}\end{eqnarray}
of the BMW $(ss,c)$ total $0.37(21)(24)\times 10^{-10}$. Considering the 
(presumably conservative) range $0\leq x_{D1}\leq 1$, 
$\left[ a_\mu^{\mathrm{sconn+disc}}\right]_{(ss,c)}$ is thus expected to 
lie somewhere between $0$ and $1/6$ times the full BMW $(ss,c)$ result, 
{\it i.e.}, in the range 
\begin{equation}
0\leq \left[ a_\mu^{\mathrm{sconn+disc}}\right]_{(ss,c)}\leq 
0.062(35)(40)\times 10^{-10}\ ,
\label{sscamusconnpdisc}\end{equation}
once more producing an essentially negligible correction to the 
strange-quark-connected plus disconnected and disconnected results,
Eqs.~(\ref{sconnpdiscnoibcorr}) and (\ref{amudiscnoibcorrinclbmw20sconn}),
above. 
\end{itemize}

If we combine the BMW-induced statistical and systematic errors in
quadrature, the $(vv,c)$, $(vv,d)$ and $(ss,c)$ contributions to the
strange-quark-connected plus disconnected sum lie in the ranges
$(-0.068(28)\times 10^{-10},0)$, $(0,0.044(14)\times 10^{-10})$, and
$(0,0.062(53)\times 10^{-10})$, respectively. Adding the errors on these 
quantities linearly, the sum of the three contributions lies between 
$-0.068(28)\times 10^{-10}$ and $0.106(67)\times 10^{-10}$, or, at 
the $1\sigma$ level, in the interval 
$(-0.096\times 10^{-10},0.173\times 10^{-10})$, leading to a conservative
final assessment of
\begin{equation}
\left[ a_\mu^{\mathrm{sconn+disc}}\right]_{(vv,c)+(vv,d)+(ss,c)}
=0.04(13)\times 10^{-10}\ .
\label{finalemamusconnpdisc}\end{equation}
The central value and error on the associated correction to the 
strange-quark-connected plus disconnected result, Eq.~(\ref{sconnpdiscnoibcorr}), 
and hence also the disconnected result, 
Eq.~(\ref{amudiscnoibcorrinclbmw20sconn}), are thus both entirely 
negligible on the scale of the errors on those results.

A similar treatment of the BMW errors and $(vv,c)$ and $(vv,d)$ 
uncertainty ranges produces an estimate for the EM contribution to 
$a_\mu^{\mathrm{MI}}$ which, at the $1\sigma$ level, lies in the interval 
$(-0.742\times 10^{-10},\, -0.245\times 10^{-10})$, leading to a 
conservative final assessment of
\begin{equation}
\left[ a_\mu^{\mathrm{MI}}\right]_{(vv,c)+(vv,d)+(ss,c)}=\, 
-0.49(25)\times 10^{-10}\ .
\label{finalemamumixed}\end{equation}
This result, like that of Eq.~(\ref{finalemamusconnpdisc}), is inclusive 
from the dispersive point of view. Unlike that earlier result, however, 
it cannot be used to remove the associated, mixed-isospin component of 
the EM contribution to the IB-uncorrected strange-quark-connected plus
disconnected result Eq.~(\ref{sconnpdiscnoibcorr}) or disconnected result
Eq.~(\ref{amudiscnoibcorrinclbmw20sconn}). The reason is that the nominally 
$I=0$ ($G$-parity negative) and $I=1$ ($G$-parity positive) exclusive-mode 
contributions to the inclusive sum in Eq.~(\ref{finalemamumixed}) enter
the difference underlying the results of Eqs.~(\ref{sconnpdiscnoibcorr}) and
(\ref{amudiscnoibcorrinclbmw20sconn}) with different signs. To correct for 
the mixed-isospin ``contamination'' of the nominal $I=0$ and $I=1$ sums, 
one thus needs to understand the breakdown of the mixed-isospin contribution 
into its exclusive-mode components. This issue is discussed in the next 
subsection.

\subsection{The mixed-isospin correction}
As is the case for the dominant isospin-conserving (IC) contribution, we 
expect IB contributions to $a_\mu^{\mathrm{LO,HVP}}$ to be dominated by 
contributions from the region of the lowest-lying ($\rho$, $\omega$) 
resonances, doubly so since IB contributions in this region are subject 
to enhancements generated by the impact on the effects of $\rho$-$\omega$ 
interference of the smallness of the $\rho$-$\omega$ mass difference. In 
this region, the mixed-isospin spectral contribution, 
$\rho_{\mathrm{EM}}^{\mathrm{MI}}(s)$, will appear essentially entirely 
in the $2\pi$ and $3\pi$ exclusive modes. 

Unlike the $ab=33,\, 88$ components, which to first order in IB receive 
only EM IB contributions, the mixed-isospin component contains both SIB 
and EM IB contributions. Eq.~(\ref{finalemamumixed}) provides an estimate 
for the latter in the EM/SIB separation convention used by BMW. Two 
determinations, one continuum ChPT based, the other lattice based, 
exist for the corresponding SIB component. The results, 
$a_\mu^{SIB}=3.32(89)\times 10^{-10}$~\cite{James:2021sor} and
$1.93(1.20)\times 10^{-10}$~\cite{Borsanyi:2020mff}, respectively, are 
compatible within errors, with a naive average of $2.83(71)\times 10^{-10}$.
The lattice result is subject to the strong cancellation between connected 
and disconnected contributions anticipated in Ref.~\cite{Lehner:2020crt}. 
Both SIB results are inclusive from the dispersive point of view. Combining, 
for example, the EM and continuum SIB estimates produces an estimate 
$2.8(9)\times 10^{-10}$ for full EM+SIB inclusive mixed-isospin contribution 
$a_\mu^{\mathrm{MI}}$. Employing instead the naive average of the continuum 
and lattice SIB results, yields the somewhat smaller value 
$2.3(8)\times 10^{-10}$.

The $\rho$-$\omega$ region $2\pi$ and $3\pi$ IB contributions to
$a_\mu^{\mathrm{LO,HVP}}$ can be estimated from the interference terms in fits
to the $e^+e^-\rightarrow 2\pi$ and $e^+e^-\rightarrow 3\pi$ electroproduction 
cross sections associated with IB $e^+e^-\rightarrow\omega\rightarrow 2\pi$ 
and $e^+e^-\rightarrow\rho\rightarrow 3\pi$ contributions to the amplitudes 
appearing in those fits. These contributions, to first order in IB, lie 
entirely in the mixed-isospin contribution, $a_\mu^{\mathrm{MI}}$.

The mixed-isospin, EM+SIB $2\pi$ contribution is taken from a fit to the 
$2\pi$ cross section based on the dispersively constrained form for the 
timelike $\pi$ form factor detailed in Ref.~\cite{Colangelo:2018mtw}. The 
parameter $\epsilon_{\omega}$ entering that form, which parametrizes
$\rho$-$\omega$ region IB, is expected to have a small non-zero 
phase~\cite{hoferichterkekgminus2meeting}. The authors of 
Ref.~\cite{Colangelo:2018mtw} have recently performed a fit to 
existing $e^+e^-\rightarrow 2\pi$ cross section data, including 
this phase as a free parameter. This fit produces a result of 
$\sim 4^o$ for the phase, and a corresponding IB $a_\mu^{\mathrm{MI}}$ 
contribution of $\sim 3.65\times 10^{-10}$~\cite{mhoferichterprivate}.
Although the fitted phase is small, the result for this contribution 
is sensitive to the inclusion of the phase in the fit, for the reason 
explained in Ref.~\cite{Wolfe:2009ts}. The same fit, with the phase 
fixed to zero by hand gives, instead, an IB $2\pi$ contribution of 
$\sim 4.32\times 10^{-10}$~\cite{mhoferichterprivate}. Given this 
sensitivity, we take the $\sim 0.67\times 10^{-10}$ difference 
between the results of the free-phase and no-phase fits as an
estimate of the uncertainty on the 
mixed-isospin, IB, $2\pi$ contribution to $a_\mu^{\mathrm{MI}}$.

The mixed-isospin, EM+SIB $\rho$-$\omega$ region $3\pi$ contribution 
is estimated using the results of VMD-based fits to recent BaBar 
$e^+ e^-\rightarrow 3\pi$ cross section data, reported in 
Ref.~\cite{BABAR:2021cde} and showing strong evidence for an IB
$\rho\rightarrow 3\pi$ interference contribution. The $a_\mu^{\mathrm{MI}}$ 
contribution produced by the interference term in the preferred 
version of this fit is $-0.56(12)\times 10^{-10}$~\cite{vdruzhininprivate}.
The error here does not account for possible additional model-dependence 
associated with the use of VMD for the IC and IB contributions to the 
amplitude. 

It is worth noting that, in spite of resonant enhancement, the magnitudes of 
the mixed-isospin, $\rho$-$\omega$ region exclusive-mode IB $2\pi$ and $3\pi$ 
contributions do not exceed the naive $\lesssim 1\% $ estimate for the size 
of IB relative to IC contributions. The sum of these contributions, 
$3.1(7)\times 10^{-10}$, is, moreover, compatible within errors with the sum 
of the results quoted above for the sum of inclusive EM and SIB contributions 
to $a_\mu^{\mathrm{MI}}$ ($2.8(9)\times 10^{-10}$ if one uses the continuum 
version of the SIB contribution, $2.3(8)\times 10^{-10}$ if one uses the 
naive average of the continuum and lattice results), confirming the 
expectation that $a_\mu^{\mathrm{MI}}$ will be dominated by $2\pi$ and 
$3\pi$ exclusive-mode contributions. The errors, however, are large enough 
to accommodate small additional contributions from the remaining, 
higher-$s$ exclusive modes. With $K\bar{K}$ and $4\pi$ contributions 
strongly dominating the sum of remaining nominal $I=0$ and $I=1$ 
contributions to $a_\mu^{\mathrm{LO,HVP}}$, we expect the magnitudes of the 
additional mixed-isospin corrections to the nominal $I=0$ and $I=1$ sums to 
be $\lesssim 1\%$ of the corresponding exclusive-mode contributions, 
{\it i.e.}, $\lesssim 0.01\, \left[ a_\mu^{\mathrm{LO,HVP}}\right]_{K\bar{K}}$ 
($\lesssim 0.36\times 10^{-10}$) and 
$\lesssim 0.01\, \left[ a_\mu^{\mathrm{LO,HVP}}\right]_{4\pi}$ 
($\lesssim 0.34\times 10^{-10}$), respectively. Even were these to
enter the correction to the nominal $a_\mu^{I=0}-{\frac{1}{9}}a_\mu^{I=1}$
combination with the same sign, the resulting correction to this
combination would be $\lesssim 0.4\times 10^{-10}$. We will thus assign
an additional $0.4\times 10^{-10}$ uncertainty to the mixed-isospin
correction to account for the missing mixed-isospin IB corrections
associated with exclusive modes other than $2\pi$ and $3\pi$.

\subsection{The final IB correction}
The estimates, Eqs.~(\ref{sconnpdiscnoibcorr}) and 
(\ref{amudiscnoibcorrinclbmw20sconn}), for the isospin-limit 
strange-quark-connected plus disconnected  contributions to 
$a_\mu^{\mathrm{LO,HVP}}$ were obtained assigning the full exclusive-mode 
$2\pi$ contribution to $a_\mu^{I=1}$ and full exclusive-mode $3\pi$ 
contribution to $a_\mu^{I=0}$. As discussed above, these exclusive-mode 
contributions contain small IB ``contaminations'' which in fact belong 
to $a_\mu^{\mathrm{MI}}$ rather than to $a_\mu^{I=1}$ or $a_\mu^{I=0}$. 
These contaminations produce an associated small mixed-isospin IB 
contamination of the nominal $a_\mu^{I=0} -{\frac{1}{9}}a_\mu^{I=1}$ 
combination obtained above. Taking the estimates just obtained for the 
mixed-isospin $2\pi$ and $3\pi$ contaminations and the bound on possible 
mixed-isospin contaminations from other exclusive modes, the mixed-isospin 
contamination to be subtracted from the result of 
Eq.~(\ref{amudiscnoibcorrinclbmw20sconn}) to obtain the true isospin-limit 
strange-quark-connected plus disconnected contribution to
$a_\mu^{\mathrm{LO,HVP}}$, is 
\begin{equation}
-0.56(12)\times 10^{-10} \, -\, {\frac{1}{9}}\left( 3.65(67)
\times 10^{-10}\right)\, \pm 0.4\times 10^{-10} \, 
=\, -0.97(14)(40)\times 10^{-10}\ ,
\label{amumi2pip3pi}\end{equation}
where the first error on the final result is the quadrature sum of 
those on the first two terms on the LHS and the second reflects
the estimate above for possible contributions from non-$2\pi$ and -$3\pi$ 
exclusive modes.

Subtracting the result (\ref{amumi2pip3pi}) from the IB-uncorrected 
nominal results $a_\mu^{\mathrm{sconn+disc}}\, =\, 39.1(1.4)\times 10^{-10}$ 
and $a_\mu^{\mathrm{disc}}\, =\, -14.3(1.4)\times 10^{-10}$ of 
Eqs.~(\ref{sconnpdiscnoibcorr}) and (\ref{amudiscnoibcorrinclbmw20sconn}), 
we obtain our final IB-corrected isospin-limit results
\begin{eqnarray}
\label{ibcorrddiscamu}
&&a_\mu^{\mathrm{sconn+disc}}\, =\, 40.1(1.4)(0.4)\times 10^{-10}\ ,\\
&&a_\mu^{\mathrm{disc}}\, =\, -13.3(1.4)(0.4)\times 10^{-10}\ .\nonumber
\end{eqnarray}

\section{An alternate analysis using results from Ref.~\cite{Davier:2019can}}
{\label{dhmz}}

A determination of exclusive-mode contributions to $a_\mu^{\mathrm{LO,HVP}}$
similar to that of Refs.~\cite{Keshavarzi:2018mgv,Keshavarzi:2019abf} 
was carried out in Ref.~\cite{Davier:2019can}, which we will refer to 
as DHMZ. The analysis above can thus be repeated using DHMZ input, 
and the results compared to those obtained using KNT2019 input. 

DHMZ's exclusive-mode results are somewhat less well suited to our 
purpose than are KNT2019's, for the following reasons. First, where 
the dispersive exclusive-mode $a_\mu^{\mathrm{LO,HVP}}$ contributions 
tabulated in KNT2019 correspond to contributions from threshold to 
$s=(1.937\ {\mathrm{GeV}})^2=3.752$~GeV$^2$, those in DHMZ correspond 
to contributions only up to $s=(1.8\ {\mathrm{GeV}})^2=3.24$~GeV$^2$. 
The pQCD approximation must thus be used to lower $s$ in a DHMZ-based 
analysis than in a KNT2019-based one. Since $R(s)$ shows a clear DV dip 
below perturbative expectations in the region between $3.24$ and 
$3.752$~GeV$^2$, this makes a DHMZ-based analyis potentially more 
sensitive to DV corrections, which can only be roughly estimated at 
present. Second, the exclusive-mode $R(s)$ contributions and covariances 
underlying the DHMZ exclusive-mode $a_\mu^{\mathrm{LO,HVP}}$ contribution 
results are not publicly available, unlike the corresponding KNT2019 
results, which are available from the authors of 
Ref.~\cite{Keshavarzi:2019abf} upon request. The access to KNT2019 
exclusive-mode data allows us to perform the internally self-consistent 
hybrid $\tau$-KNT2019-electroproduction determination of the $K\bar{K}$ 
contribution to $a_\mu^{\mathrm{sconn+disc}}$ described above. The lack 
of access to the corresponding DHMZ exclusive-mode cross sections and 
covariances means an analogous, fully self-consistent DHMZ-based 
determination of that contribution is not possible. We have thus been 
forced to use a KNT2019-based determination of the $I=1$ $K\bar{K}$ 
contribution from the region between $2.7556$ GeV$^2$ and $3.24$ GeV$^2$ 
to determine the full hybrid $\tau$-electroproduction $I=1$ 
DHMZ-exclusive-mode-region $K\bar{K}$ contribution, combining that with 
the DHMZ result for the full $I=0+1$ $K\bar{K}$ contribution up to 
$3.24$ GeV$^2$, to determine the ``DHMZ-based'' $K\bar{K}$ contribution 
to $a_\mu^{\mathrm{sconn+disc}}$.

While use of DHMZ input has some minor disadvantages, it is of 
interest to pursue the alternate, DHMZ-based determination of
$a_\mu^{\mathrm{sconn+disc}}$ and $a_\mu^{\mathrm{disc}}$ since KNT2019 
and DHMZ, despite analyzing essentially identical electroproduction 
cross-section data, obtain somewhat discrepant results for a 
number of exclusive-mode $a_\mu^{\mathrm{LO,HVP}}$ contributions 
(the situation is discussed in more detail in Sec.~2.3.5 of 
Ref.~\cite{Aoyama:2020ynm}, with Table 5 providing a summary of the 
main discrepancies). Such discrepancies exist for both nominally $I=0$ 
and nominally $I=1$ exclusive-mode contributions and have the potential 
to affect the weighted difference of $I=0$ and $I=1$ contributions 
which determines $a_\mu^{\mathrm{sconn+disc}}$.

The implementation of the DHMZ-based analysis follows exactly that of
the KNT2019-based analysis, detailed above. DHMZ results for the various 
exclusive-mode $a_\mu^{\mathrm{LO,HVP}}$ contributions are taken from 
Table 2 of DHMZ.{\footnote{The reader is reminded that results 
tabulated in KNT2019 and DHMZ, though having the same exclusive-mode 
labellings, are different, corresponding to contributions from different 
ranges of $s$ (for DHMZ, up to $s=3.24$ GeV$^2$, for KNT2019, up to 
$s=3.752$ GeV$^2$), and thus should not be compared mode by mode. A 
comparison of contributions for a subset of exclusive modes over the 
common, DHMZ range $s\leq 3.24$ GeV$^2$ is provided in Table 5 of 
Ref.~\cite{Aoyama:2020ynm}. The lower DHMZ upper endpoint, which lies 
below $p\bar{p}$ threshold, also means there are no DHMZ analogues of 
the KNT2019 entries for the $G$-parity-ambiguous $p\bar{p}$ and 
$n\bar{n}$ mode contributions.}} 
Conversions of $G$-parity-ambiguous-mode $a_\mu^{\mathrm{LO,HVP}}$
contributions to the corresponding $a_\mu^{\mathrm{sconn+disc}}$ 
contributions proceed exactly as in the case of the KNT2019-based 
analysis, with the exception of the $K\bar{K}$ mode, where, as noted 
above, we do not have access to the DHMZ $K\bar{K}$ exclusive-mode 
cross sections and covariances, and hence have used KNT2019 results 
for these quantities to determine the $K\bar{K}$ contribution to 
$a_\mu^{\mathrm{LO,HVP}}$ from the region $2.7556$ GeV$^2$
$\leq s\leq 3.24$ GeV$^2$. 

The results of the DHMZ-based analysis are as follows. 

The sums of the nominally $I=1$ ($G$-parity positive) and $I=0$ 
($G$-parity negative) exclusive-mode contributions from Table 2 
of DHMZ are $542.74(84)(3.28)(1.12)\times 10^{-10}$ and 
$53.65(42)(1.11)(1.02)\times 10^{-10}$, respectively, where the errors
are, in order, the statistical, mode-specific mode-to-mode-uncorrelated 
systematic, and $100\%$ correlated common systematic error components of 
Ref.~\cite{Davier:2019can}. Combining the first two errors in quadrature, 
these results become $542.74(3.39)(1.12)\times 10^{-10}$ and 
$53.65(1.18)(1.02)\times 10^{-10}$, respectively.

For the $K\bar{K}$ mode, the $I=1$ contribution to $a_\mu^{\mathrm{LO,HVP}}$ 
from the region $s\leq 2.7556$ GeV$^2$, is obtained, as before, using BaBar 
$\tau$ data~\cite{BaBar:2018qry}. The result, $0.764(33)\times 10^{-10}$, 
is thus unchanged. KNT2019 input produces a ``maximally conservative''
estimate of $0.070(70)\times 10^{-10}$ for the $I=1$ contribution from the
region $2.7556$ GeV$^2$ $\leq s\leq 3.24$ GeV$^2$. The resulting $s\leq 3.24$ 
GeV$^2$, $I=1$ contribution, $0.834(34)\times 10^{-10}$, combined with the 
DHMZ two-mode $K\bar{K}$ $I=0+1$ contribution total, then yields 
$\left[ a_\mu^{\mathrm{sconn+disc}}\right]_{K\bar{K}}=34.98(43)(36)$,
where the second error is the (linear) sum of the $100\%$ correlated 
common systematic DHMZ errors on the $K^+K^-$ and $K_SK_L$ contributions.

For the $K\bar{K}\pi$ mode, using the results for the $I=1$ component
of the observed cross sections obtained by BaBar~\cite{BaBar:2007ceh}, 
one finds an $I=1$ contribution to $a_\mu^{\mathrm{LO,HVP}}$ from the 
region $s\leq 3.24$ GeV$^2$ of $0.664(34)(105)\times 10^{-10}$,
where the first error is statistical and second systematic. Combining
these errors in quadrature, the DHMZ result for the full $I=0+1$ 
contribution then implies 
$\left[ a_\mu^{\mathrm{sconn+disc}}\right]_{K\bar{K}\pi}=1.71(17)(6)$,
where the second error is again the $100\%$ correlated common systematic 
error on the DHMZ $K\bar{K}\pi$ total.

For the $K\bar{K}2\pi$ mode, the $s\leq 3.24$ GeV$^2$, $I=0$ 
$\phi (\rightarrow K\bar{K})\pi\pi$ contribution implied by BaBar 
$e+e^-\rightarrow \phi \pi\pi$ cross-section results~\cite{BaBar:2011btv} 
is $0.117(8)\times 10^{-10}$. Subtracting this from the DHMZ $I=0+1$ 
$K\bar{K}2\pi$ total and performing the usual ``maximally conservative'' 
treatment of the resulting $G$-parity-ambiguous residual, one finds 
$\left[ a_\mu^{\mathrm{sconn+disc}}\right]_{K\bar{K}2\pi}=0.44(41)(0)$,
where the second error is, once more, the $100\%$ correlated common 
systematic DHMZ one.

For the remaining ($\omega K\bar{K}$ and 
$\eta K\bar{K}\, ({\mathrm{no}}\ \phi )$) DHMZ 
$G$-parity-ambiguous modes, the ``maximally conservative'' treatment of the 
sum of contributions from these modes yields a contribution of
$0.00(1)(0)\times 10^{-10}$ to $a_\mu^{\mathrm{sconn+disc}}$, with the
second error again the $100\%$ correlatated common systematic DHMZ one.

Finally, the five-loop-truncated pQCD estimate for the contribution to 
$a_\mu^{\mathrm{sconn+disc}}$ from the DHMZ inclusive region,
$s> 3.24$ GeV$^2$ is found to be $7.28\times 10^{-10}$,
with, as above, negligible input-$\alpha_s$ and five-loop-truncation 
uncertainties.

Combining these results we find, for our DHMZ-based determination
of the (pre-IB-corrected) strange-quark-connected plus disconnected sum
\begin{eqnarray}
\label{sconnpdiscnoibcorrDHMZ}
a_\mu^{\mathrm{sconn+disc}}&=&\biggl[ 53.65(1.18)(1.02)_{\mathrm{lin}}-
{\frac{542.73(3.39)(1.12)_{\mathrm{lin}}}{9}} +34.98(43)(36)_{\mathrm{lin}}\\
&&\phantom{\biggl[}+1.71(17)(6)_{\mathrm{lin}} 
+0.44(41)(1)_{\mathrm{lin}}
+0.00(1)(0)_{\mathrm{lin}}+7.28\biggr]\times 10^{-10}\nonumber\\
&=&37.76(1.39)(1.33)_{\mathrm{lin}}\times 10^{-10}\ ,\nonumber
\end{eqnarray}
where the $100\%$ correlated common systematic errors are identified
by the subscript ``$\mathrm{lin}$''. The first error in the final expression
in Eq.~(\ref{sconnpdiscnoibcorrDHMZ}) is the quadrature sum of the 
statistical and uncorrelated mode-specific systematic errors, the 
second the linear sum of the $100\%$ correlated common systematic 
errors. This treatment of the common systematic errors is that
specified in Ref.~\cite{Davier:2019can}.

Comparing Eqs.~(\ref{sconnpdiscnoibcorr}) and (\ref{sconnpdiscnoibcorrDHMZ}),
we see that (a) the DHMZ-based value is $1.32\times 10^{-10}$ lower 
than the KNT2019-based value; (b) the DHMZ-based total error is more 
conservative.

Using the naive average of the whitepaper strange-quark-connected 
result~\cite{Aoyama:2020ynm} and the BMW strange-quark-connected 
result~\cite{Borsanyi:2020mff}, we find for our initial 
(pre-IB-corrected) DHMZ-based disconnected contribution the value 
$-15.61(1.39)(1.33)_{\mathrm{lin}}\times 10^{-10}$, to be 
compared with Eq.~(\ref{amudiscnoibcorrinclbmw20sconn}).

Finally, applying the IB corrections worked out in Sec.~\ref{ibcorrns},
we arrive at
\begin{eqnarray}
\label{ibcorrddiscamuDHMZ}
&&a_\mu^{\mathrm{sconn+disc}}\, =\, 38.7(1.4)(1.3)_{\mathrm{lin}}(0.4)
\times 10^{-10}\ ,\\
&&a_\mu^{\mathrm{disc}}\, =\, -14.6(1.4)(1.3)_{\mathrm{lin}}(0.4)
\times 10^{-10}\ ,\nonumber
\end{eqnarray}
where the third error has the same origin as the second error in
Eq.~(\ref{ibcorrddiscamu}).

Table~\ref{tablattdisc} compares our isospin-symmetric results for 
$a_\mu^{\mathrm{sconn+disc}}$ and $a_\mu^{\mathrm{disc}}$ with those of 
recent lattice studies that report results for both. Our disconnected 
results are seen to be in excellent agreement with all lattice results
except that of the Mainz collaboration~\cite{Gerardin:2019rua},{\footnote{The 
result of Ref.~\cite{Gerardin:2019rua} was obtained via an extrapolation to the
physical point from results at heavier-than-physical pion masses, unlike 
the results obtained by the other collaborations.}} with which they are
clearly incompatible.

\begin{table}[h]
\begin{center}
\begin{tabular}{llll}
\hline
Source&$\ n_f\ $&$\quad a_\mu^{\mathrm{disc}}\times 10^{10}$
&$\quad a_\mu^{\mathrm{sconn+disc}}\times 10^{10}$\\
\hline
RBC/UKQCD~\cite{RBC:2015you,RBC:2018dos}&$2+1$&$\quad -11.2(3.3)(2.3)$
&$\quad 42.0(3.3)(2.3)$\\
BMW~\cite{Budapest-Marseille-Wuppertal:2017okr}&$2+1+1$&$\quad 
-12.8(1.1)(1.6)$&$\quad 40.9(1.2)(1.7)$\\
Mainz~\cite{Gerardin:2019rua}&$2+1$&$\quad -23.2(2.2)(4.5)$
&$\quad 31.3(3.3)(4.5)$\\
BMW~\cite{Borsanyi:2020mff}&$2+1+1$&$\quad -13.36(1.18)(1.36)$
&$\quad 40.03(1.18)(1.36)$\\
\hline
This work, Eqs.~(\ref{ibcorrddiscamu})&$2+1$&$\quad -13.3(1.4)(0.4)$
&$\quad 40.1(1.4)(0.4)$\\
This work, Eqs.~(\ref{ibcorrddiscamuDHMZ})&$2+1$&
$\quad -14.6(1.4)(1.3)_{\mathrm{lin}}(0.4)$&
$\quad 38.7(1.4)(1.3)_{\mathrm{lin}}(0.4)$\\
\hline
\end{tabular}
\caption{\label{tablattdisc} {\it {Comparison of our results with
recent lattice results for the isospin-limit three-flavor 
disconnected and strange-quark-connected plus full, three-flavor
disconnected contributions to $a_\mu^{\mathrm{LO,HVP}}$. First 
and second errors on lattice entries, shown in the upper half of the 
table, are statistical and systematic, respectively. Errors on the 
KNT2019- and DHMZ-based results of Eqs.~(\ref{ibcorrddiscamu}) and 
(\ref{ibcorrddiscamuDHMZ}), shown in the second last and last
lines of the lower half of the table, respectively, are as described 
in the text.}}}
\end{center}
\end{table}

The errors on the results for $a_\mu^{\mathrm disc}$ in 
Eqs.~(\ref{ibcorrddiscamu}) and\ (\ref{ibcorrddiscamuDHMZ}) are 
competitive with those of the most precise of the current lattice 
results~\cite{Borsanyi:2020mff}, and obtained with dramatically reduced 
numerical cost.

\section{Discussion}\label{discussion}
The main observations of this paper are (i) that a rather precise 
determination of $a_\mu^{\mathrm{sconn+disc}}$, the strange-quark-connected 
plus full, three-flavor quark-disconnected contribution to 
$a_\mu^{\mathrm{LO,HVP}}$ can be obtained from electroproduction data and 
(ii) that, using lattice results for the strange-quark-connected 
contribution, this can be converted into a determination of the 
disconnected part, avoiding the direct calculation of the disconnected 
part on the lattice. While a completely lattice-based evaluation of 
$a_\mu^{\mathrm{LO,HVP}}$ is of great interest, this disconnected part 
is computationally expensive, and a ``hybrid'' approach, in which the 
disconnected part is obtained from data, and the connected part from 
the lattice, is of interest as well. Moreover, as already mentioned 
in Sec.~\ref{intro}, it is useful to compare results for different 
contributions to $a_\mu^{\mathrm{LO,HVP}}$ obtained using dispersive 
and lattice approaches.

It is worth reminding the reader that, to make contact with the 
disconnected contributions calculated on the lattice, our results for 
$a_\mu^{\mathrm{disc}}$ are isospin-symmetric ones. This means the EM 
disconnected contributions shown in Fig.~\ref{emibgraphs} still have 
to be added to obtain a complete result for $a_\mu^{\mathrm{LO,HVP}}$.
Since these corrections are much smaller than the isospin-symmetric 
disconnected contribution, however, it suffices, at a given level of 
overall precision, to evaluate them on the lattice with much larger 
relative errors.

While one of our two main observations is that the full, three-flavor,
isospin-symmetric disconnected contribution to $a_\mu^{\mathrm{LO,HVP}}$ 
can be obtained without the need for any disconnected lattice calculations, 
we did use some disconnected results (notably, the BMW results for diagrams 
F and D1 in Fig.~\ref{emibgraphs}) in our discussions of IB corrections in 
Sec.~\ref{ibcorrns}. We stress, however, that although BMW input was used, 
the specifics of that input were not, in fact, numerically relevant, for 
the following reasons. First, the BMW EM results were not needed for the 
mixed-isospin EM corrections, where (i) the dominant combined SIB+EM 
$2\pi$ and $3\pi$ exclusive-mode contributions from the $\rho$-$\omega$ 
region were taken from experiment and (ii) since (in spite of the 
enhancement of these IB contributions from the rather small $\rho$-$\omega$ 
mass difference) these contributions are each $\sim 1\%$ of the total IC 
contributions from these modes, it should be quite safe to estimate 
mixed-isospin corrections from other exclusive modes using the $\sim 1\%$ 
estimate for each mode and adding these linearly. The mixed-isospin 
EM+SIB correction is thus under control, without needing any lattice 
disconnected input (EM or otherwise), up to the $\sim 0.4\times 10^{-10}$ 
uncertainty for non-$2\pi ,\, 3\pi$ exclusive-mode contributions.
Second, the general analysis of the EM corrections to the $ab=33$ and
$88$ parts of $a_\mu^{\mathrm{LO,HVP}}$ shows these involve very 
strong cancellations. From the general forms, written in terms of 
the $SU(3)_F$-breaking loop factor ratios, it is clear these strong 
cancellations are entirely generic. Thus, unless the EM corrections 
to these quantities are much larger than the scale of the BMW results, 
we can be sure these corrections are small without needing to know the 
precise values of the valence-valence disconnected EM contribution or 
the unsuppressed sea-sea connected contribution from the graph $D1$.
Although some disconnected EM lattice results have been used in the
discussion above, the only substantive role these play is to confirm 
that the EM disconnected contributions do not have a massive enhancement 
relative to the expected ``natural'' $\sim 1\%$ EM IB scale. 
\vskip .15in
To conclude, we return to the reliability of treating the inclusive 
region (above $s=(1.937\ {\mathrm{GeV}})^2$ for KNT2019, above 
$s=(1.8\ {\mathrm{GeV}})^2$ for DHMZ) using pQCD. To be specific, we  
focus our discussion on the KNT2019 case. Uncertainties due to 
truncation in order and the uncertainty in the input $\alpha_s$ used 
are tiny, as are perturbative $D=2$ corrections (the latter for the 
reasons outlined in Sec.~\ref{pcont}). The main uncertainty associated 
with the use of pQCD for the contribution from this region will thus 
most likely be that due to residual duality-violating (DV) corrections. 
Let us consider the {\it ansatz} for the EM DVs used in the determination 
of $\alpha_s$ from electroproduction cross-section data, detailed in 
Ref.~\cite{Boito:2018yvl}. This has the form
\begin{equation}
\rho_{\mathrm{EM}}^{\mathrm{DV}}(s)={\frac{5}{9}}\rho_{ud;V}^{\mathrm{DV}}(s)
+{\frac{1}{9}}\rho_{0,V}^{\mathrm{DV}}(s)\ ,
\label{emalphaspaperdvansatz}\end{equation}
where $\rho_{ud;V}^{\mathrm{DV}}(s)$ is taken to have the large-$s$ form,
\begin{equation}
\rho_{ud;V}^{\mathrm{DV}}(s)=\mbox{exp}\left( -\delta_1 -\gamma_1 s\right)\,
\sin\left( \alpha_1+\beta_1 s\right)\ ,
\label{lgencreggedvform}\end{equation}
characterized by the DV parameters $\delta_1$, $\gamma_1$, $\alpha_1$
and $\beta_1$, which follows for massless quarks from the large-$N_c$ 
and Regge arguments discussed in Ref.~\cite{Boito:2017cnp}. These DV
parameters can be obtained from finite-energy-sum-rule (FESR) fits to 
weighted integrals of the $I=1$ vector-current spectral distributions
measured in non-strange hadronic $\tau$ decays. The $\tau$-based results
and their covariances were used as priors in the EM fits of
Ref.~\cite{Boito:2018yvl}. The strange-quark contribution, 
$\rho_{0,V}^{\mathrm{DV}}(s)$ was taken to have the same functional form, 
though with generally different DV parameters, reflecting the shift 
of resonances with hidden strangeness to higher locations in the spectrum. 
Ref.~\cite{Boito:2017cnp} took $\gamma_0= \gamma_1$ and $\beta_0= \beta_1$, 
and fitted $\alpha_0$ and $\delta_0$ as free parameters, simultaneously 
re-fitting $\delta_1$, $\gamma_1$, $\alpha_1$ and $\beta_1$, subject to 
the input $\tau$-based prior constraints. Let us rewrite the 
{\it ansatz}~(\ref{emalphaspaperdvansatz}) in the alternate form with 
$I=0$ and $1$ contributions explicitly separated:
\begin{equation}
\rho_{\mathrm{EM}}^{\mathrm{DV}}(s)={\frac{1}{2}}\rho_{ud;V}^{\mathrm{DV}}(s)
+\left[ {\frac{1}{18}}\rho_{ud;V}^{\mathrm{DV}}(s)
+ {\frac{1}{9}}\rho_{0,V}^{\mathrm{DV}}(s)\right] \ .
\label{emalphaspaperdvalt}\end{equation}
The term in the square brackets is the $I=0$ contribution. In this form, it 
is obvious that the DV {\it ansatz} of Ref.~\cite{Boito:2018yvl} produces a 
DV contribution to the $\rho_{\mathrm{EM}}^{\mathrm{sconn+disc}}(s)$ 
combination in which the light-quark part exactly cancels, leaving 
\begin{equation}
\rho_{\mathrm{EM}}^{\mathrm{sconn+disc};\mathrm{DV}}(s)=
{\frac{1}{9}}\rho_{0,V}^{\mathrm{DV}}(s)\ .
\label{rhosconnpdiscdv}\end{equation}
In view of the significant recent improvement in the FESR analysis
of the $I=1$ vector channel reported in Ref.~\cite{Boito:2020xli},
we have updated the EM fits of Ref.~\cite{Boito:2018yvl} using as new 
light-quark DV priors the improved versions obtained in the fits of 
Ref.~\cite{Boito:2020xli}. Using a range of different choices for the 
$\tau$ and $\mathrm{EM}$ FESR fit windows, we find integrated DV 
contributions to $a_\mu^{\mathrm{LO,HVP}}$ from the region above 
$s=(1.937\ {\mathrm GeV})^2$ lying in a narrow range around 
$-0.25\times 10^{-10}$. In view of the fact that the functional form 
for the DV ansatz in Eq.~(\ref{lgencreggedvform}) was derived only in 
the massless limit, but has been employed also for the strange-quark 
contribution, we treat this result as providing only a rough estimate of 
the uncertainty associated with neglecting possible residual integrated 
DV contributions in the region above $s=(1.937\ {\mathrm ~GeV})^2$. Even 
were one to double this estimate, however, the resulting 
$\sim 0.5\times 10^{-10}$ DV-induced uncertainty would still be small 
on the scale of the $\sim 1.5\times 10^{-10}$ uncertainty of the 
strange-quark-connected plus disconnected and disconnected results 
obtained above. 
 
\begin{acknowledgments}
We thank Vladimir Druzhinin of BaBar for the providing the
$\rho\rightarrow 3\pi$ IB interference contribution portion of 
the $3\pi$ contribution to $a_\mu^{\mathrm{LO,HVP}}$ corresponding 
to the fit to $e^+ e^-\rightarrow 3\pi$ cross section described in 
Ref.~\cite{BABAR:2021cde}, Gilberto Colangelo, Martin Hoferichter 
and Peter Stoffer for providing the results of the zero-phase and 
free-phase dispersive fit determinations of the mixed-isospin IB 
$2\pi$ exclusive-mode contribution from the $\rho$-$\omega$ 
interference region, and Alex Keshavarzi for providing details of 
the combined KNT2019 exclusive-mode cross-section results 
used in the analysis above. DB is supported by the S\~ao Paulo 
Research Foundation (FAPESP) Grant No.~2021/06756-6, by
CNPq Grant No.~309847/2018-4,and by Coordena\c c\~ao de 
Aperfei\c coamento de Pessoal de N\'ivel Superior -- Brasil (CAPES) -- 
Finance Code 001. MG is supported by the U.S.\ Department 
of Energy, Office of Science, Office of High Energy Physics, under 
Award No. DE-SC0013682. KM is supported by a grant from the Natural 
Sciences and Engineering Research Council of Canada. SP is supported 
by the Spanish Ministry of Science, Innovation and Universities 
(project PID2020-112965GB-I00/AEI/10.13039/501100011033) and by 
Grant No. 2017 SGR 1069. IFAE is partially funded by the CERCA 
program of the Generalitat de Catalunya. DB thanks the University
of Vienna for hospitality.
\end{acknowledgments}
\vskip1cm
\appendix*
\section{Remaining channels}

The ``maximally conservative'' assessments for the remaining 
$G$-parity-ambiguous exclusive-mode contributions from the KNT2019 
list are as follows:
\begin{itemize}
\item[$\circ$] $K\bar{K} 2\pi$: KNT2019 gives an $I=0+1$ total of 
$[a_\mu^{\mathrm{LO,HVP}}]_{K\bar{K}2\pi} = 1.93(8) \times 10^{-10}$. 
Part of this contribution comes from the $G$-parity negative, $I=0$ 
mode $\phi\pi\pi$. The BaBar $e^+e^-\rightarrow \phi\pi\pi$ cross section
results~\cite{BaBar:2011btv} (which were obtained by dividing the 
observed $\phi$-region $K^+ K^- \pi^+\pi^-$ cross sections by the 
then-current, 2010 PDG $\phi\rightarrow K^+ K^-$ branching fraction) 
produce a corresponding $I=0$ all-$\phi$-decay-mode $\phi\pi\pi$ 
contribution to $a_\mu^{\mathrm{LO,HVP}}$ of $0.192(12)\times 10^{-10}$. 
The $\phi (\rightarrow 3\pi )\pi\pi$ part of this contribution is already 
included in the listed $5\pi$ contributions. We have updated 
the BaBar-based all-$\phi$-decay-mode result quoted above using the 
current PDG result for the $\phi\rightarrow K^+ K^-$ branching 
fraction~\cite{ParticleDataGroup:2020ssz}. Multiplying this result
by the current PDG two-mode $\phi\rightarrow K\bar{K}$ branching
fraction sum, one finds the result $0.159(10)\times 10^{-10}$ for the 
$I=0$, $\phi (\rightarrow K\bar{K})\pi\pi$ contribution to 
$[a_\mu^{\mathrm{LO,HVP}}]_{K\bar{K}2\pi}$. Subtracting this from the 
total $K\bar{K}2\pi$ contribution leaves a residual $G$-parity-ambiguous 
$K\bar{K} 2\pi$ contribution of $1.77(8)\times 10^{-10}$. Combining the 
small $I=0$, $\phi (\rightarrow K\bar{K})\pi\pi$ contribution with the 
generic ``maximally conservative'' treatment of the latter then gives
a strange-quark-connected plus disconnected contribution of
\begin{equation}
\left[ a_\mu^{\mathrm{sconn+disc}}\right]_{K\bar{K}2\pi}
= 0.95(98)\times 10^{-10}\, .
\label{sconnpdisckkbar2pi}\end{equation}

\item[$\circ$] $K\bar{K} 3\pi$: KNT2019 gives an $I=0+1$ total of 
$[a_\mu^{\mathrm{LO,HVP}}]_{K\bar{K}3\pi} = 0.04(2) \times 10^{-10}$. 
The generic ``maximally conservative'' bound treatment thus gives
a strange-quark-connected plus disconnected contribution of
\begin{equation}
[a_\mu^{\mathrm{sconn+disc}}]_{K\bar{K}3\pi}= 0.02(2)\times 10^{-10}\, .
\label{sconnpdisckkbar3pi}\end{equation}

\item[$\circ$] $X_1=\omega(\rightarrow npp)K\bar{K}$ and
$X_2=\eta (\rightarrow npp) K\bar{K}$ (no $\phi$): KNT2019 gives
$I=0+1$ totals of $[a_\mu^{\mathrm{LO,HVP}}]_{X_1}=0.00(0)\times 10^{-10}$ 
and $[a_\mu^{\mathrm{LO,HVP}}]_{X_2}=0.01(1)\times 10^{-10}$ for these 
two modes. The generic ``maximally conservative'' bound treatment thus 
gives a strange-quark-connected plus disconnected contribution of 
\begin{equation}
[a_\mu^{\mathrm{sconn+disc}}]_{X_1+X_2}= 0.00(1)\times 10^{-10}\, 
\label{sconnpdiscx1x2}\end{equation}
for the sum of the contributions from these two modes.
\item[$\circ$] $p\bar{p}$ and $n\bar{n}$: KNT2019 gives $I=0+1$ totals of 
$[a_\mu^{\mathrm{LO,HVP}}]_{p\bar{p}}=0.03(0)\times 10^{-10}$ and 
$[a_\mu^{\mathrm{LO,HVP}}]_{n\bar{n}}=0.03(1)\times 10^{-10}$ for these 
two modes. The generic ``maximally conservative'' bound treatment thus 
gives a strange-quark-connected plus disconnected contribution of 
\begin{equation}
[a_\mu^{\mathrm{sconn+disc}}]_{p\bar{p}+n\bar{n}}= 0.03(3)\times 10^{-10}\, 
\label{sconnpdiscNNbar}\end{equation}
for the sum of the contributions from these two modes.
\item[$\circ$] Low-$s$ $\pi^0 \gamma$ and $\eta \gamma$: The 
$\pi^0 \gamma$ and $\eta \gamma$ contributions from the higher-$s$ 
region are strongly dominated by contributions from the large $\omega$ 
and $\phi$ peaks in the experimental cross sections, and hence 
identifiable as $I=0$. Such an $I=0$ assignment is, however, less 
certain at low-$s$, and we thus use the ``maximally conservative'' 
treatment in this region. With KNT2019 giving low-$s$ $I=0+1$ totals of 
$0.12(1)\times 10^{-10}$ and $0.00(0)\times 10^{-10}$ for these
modes, this leads to a strange-quark-connected plus disconnected 
contribution of
\begin{equation}
[a_\mu^{\mathrm{sconn+disc}}]_{low-s\ \pi^0\gamma +\eta\gamma}= 
0.05(7)\times 10^{-10}\, 
\label{lowspi0gammaetagamma}\end{equation}
for the sum of the contributions from these two modes.
\end{itemize}

\vfill\eject
\end{document}